\documentclass[aps,prd,nofootinbib,showpacs,preprintnumbers]{revtex4-1}
\usepackage{epsfig}
\usepackage[dvips]{color}
\usepackage{amsmath}

\usepackage{soul}
\setstcolor{red}
\sethlcolor{yellow}

\def\lsim{\raise0.3ex\hbox{$<$\kern-0.75em\raise-1.1ex\hbox{$\sim$}}}
\def\gsim{\raise0.3ex\hbox{$>$\kern-0.75em\raise-1.1ex\hbox{$\sim$}}}

\makeatletter
\@addtoreset{equation}{section}
\makeatother

\newcommand{\beqn}{\begin{equation}}
\newcommand{\eqn}{\end{equation}}
\newcommand{\bqa}{\begin{eqnarray}}
\newcommand{\eqa}{\end{eqnarray}}
\newcommand{\bqas}{\begin{eqnarray*}}
\newcommand{\eqas}{\end{eqnarray*}}
\newcommand{\bdm}{\begin{displaymath}}
\newcommand{\edm}{\end{displaymath}}

\begin{document}
\thispagestyle{empty}
 \mbox{} \hfill BI-TP 2010/46\\
\begin{center}
{{\large \bf Thermal dilepton rate and electrical conductivity: \\[1mm]
An analysis of vector current correlation functions in \\[1mm]
quenched lattice QCD} 
 } \\
\vspace*{1.0cm}
{\large H.-T. Ding$^{1,2}$, A. Francis$^1$, O. Kaczmarek$^1$, 
F. Karsch$^{1,2}$, \\
E. Laermann$^1$ and W. Soeldner$^{1}$}

\vspace*{1.0cm}

{\normalsize
$^{\rm 1}$ Fakult\"at f\"ur Physik, Universit\"at Bielefeld, D-33615 Bielefeld, Germany\\
$^{\rm 2}$ Physics Department, Brookhaven National Laboratory, Upton, NY 11973, USA \\
}
\end{center}
\vspace*{1.0cm}
\centerline{\large ABSTRACT}

\baselineskip 20pt

\noindent
We calculate the vector current correlation function for light 
valence quarks in the deconfined phase of QCD. The calculations
have been performed in quenched lattice QCD at $T\simeq 1.45 T_c$ for 
four values of the lattice cut-off on lattices up to size $128^3\times 48$.
This allows to perform a continuum extrapolation of the correlation
function in the Euclidean time interval 
$0.2\le \tau T \le 0.5$,
which extends to the largest temporal separations possible at finite
temperature, to
better than 1\% accuracy.
In this interval, at the value of the temperature investigated,
we find 
that the vector correlation function never deviates
from the free correlator for massless quarks by more than 9\%.
We also determine the first two non-vanishing thermal moments of the vector 
meson spectral function. 
The second thermal moment deviates by less than 7\% from the free
value.
With these constraints, we then proceed to extract information on the
spectral representation of the vector correlator and
discuss resulting consequences for the electrical conductivity 
and the thermal dilepton rate in 
the plasma phase.
\vfill
\eject
\baselineskip 15pt

\section{Introduction}

Many of the properties of deconfined, strongly interacting matter
are reflected in the structure of correlation functions of the
vector current. Its spectral representation directly relates to
the invariant mass spectrum of dileptons and photons.
Depending on the invariant mass regime one can get information
on in-medium properties of heavy \cite{Satz} and light \cite{light} 
quark bound states. For low invariant masses one enters the hydrodynamic
regime where spectral functions reflect transport properties of the
thermal medium \cite{Boon,Forster,Aarts02,Moore,Teaney06,Teaney10}. 
In the case of the 
vector correlation function and its spectral representation this is 
characterized by a transport coefficient,
the electrical conductivity. Heavy ion experiments start getting sensitive
to the relevant regime of low mass dileptons/photons of a few hundred MeV, 
{\it i.e.} invariant masses that are of the order of the scale set by the 
QCD transition temperature \cite{Phenix}.

Spectral functions in the vector channel, or equivalently cross sections
for thermal dilepton production have been extensively studied in perturbation
theory \cite{PTtheory,Blaizot}. Non-perturbative aspects have been included
through the hard thermal loop resummation scheme \cite{Braaten}. A
straightforward application of the latter,
however, breaks down in the low invariant mass regime. This finds its 
manifestation in a 
too rapid rise of the spectral density which, in turn, leads to an  
infrared divergent Euclidean correlator of the vector current \cite{Thoma}. 
This low invariant mass regime has been examined in much detail \cite{Arnold,Gelis}
and it has been demonstrated that the spectral function at low energies
will increase linearly with energy giving rise to a finite, non vanishing  
electrical conductivity of the quark gluon plasma. 

Lattice calculations of Euclidean time correlation functions can give
access to spectral functions \cite{Wyld}. Extracting these directly,
however, is an ill-posed problem and progress in this direction has been 
made only recently through the application of statistical tools like the
Maximum Entropy Method (MEM) \cite{Asakawa}.  
Subsequently this method has been applied to also analyze
correlation functions of the vector current \cite{our_dilepton,Gupta,Aarts07}.

In this paper we
analyze the behavior of the vector correlation function 
at high temperature, {\it i.e.} in the deconfined phase of QCD. 
Our first
goal is to generate precise data for this correlator which 
allow for its extrapolation to the continuum limit in a time interval
extending to the largest temporal separations possible at finite $T$.
From the large $\tau$ behavior of the correlator we then collect
evidence for non-perturbative structure of the corresponding
spectral function at low energy. This forms the basis for the extraction
of information on the spectral representation of the vector correlator
by starting with an ansatz for the spectral function as well as
by exploiting the Maximum Entropy Method (MEM).

The paper is organized as follows. In the next section we collect basic 
formulas for vector correlation functions and their spectral representation
which are related
to the thermal dilepton rate as well as the
electrical conductivity. Section III introduces thermal moments of vector
correlation functions. In Section IV we present our numerical results 
on the vector correlation function calculated at $T\simeq 1.45T_c$. In 
Section V we discuss an analysis of these results based on fits to the data 
as well as on an analysis using the MEM. In
Section VI we discuss consequences for the calculation of dilepton and photon
rates as well as the electrical conductivity. 
We finally give our conclusions and an outlook on future studies
in Section VII.

\section{Thermal vector correlation function, dilepton rate and
electrical conductivity}

\subsection{Correlation functions}

Let us start by summarizing some basic relations between the Euclidean 
correlation function of the vector current,
\begin{equation}
J_\mu (\tau,\vec{x}) \equiv \bar{q}(\tau, \vec{x})\gamma_\mu q(\tau, \vec{x})\; ,
\label{V_current}
\end{equation}
at non-zero temperature and its 
spectral representation on the one hand and thermal dilepton rates and the 
electrical conductivity on the other hand. 

We will analyze  Euclidean time two-point functions of the current $J_\mu$
at fixed momentum,
\begin{equation}
G_{\mu\nu}(\tau,\vec{p}) = \int {\rm d}^3x\ G_{\mu\nu}(\tau,\vec{x}) \
{\rm e}^{i \vec{p} \cdot \vec{x}} \; ,
\label{corr_p}
\end{equation}
with
\begin{equation}
G_{\mu\nu}(\tau,\vec{x}) =
\langle J_\mu (\tau, \vec{x}) J_\nu^{\dagger} (0, \vec{0}) \rangle \; .
\label{tempcor}
\end{equation}
Here $\langle ...\rangle$ denotes thermal expectation values.

The vector current receives contributions from connected as well as 
disconnected diagrams. However, contributions from the latter are 
parametrically small. In high temperature perturbation theory they occur 
at ${\cal O}(g^6\ln 1/g)$ and give rise to the difference 
between quark number and isospin susceptibilities \cite{Blaizotsus}.
This difference has been found to be small at all temperatures in the
high temperature phase of QCD \cite{GG,Allton6}. 
Moreover, in calculations of the correlation function
of the electromagnetic current,
\begin{equation}
J_{em} = \sum_f Q_f \bar{q}^f(\tau, \vec{x})\gamma_\mu q^f(\tau, \vec{x})
\; ,
\label{Jem}
\end{equation}
the contribution of disconnected diagrams vanishes for three degenerate
quark flavors as it is proportional to the square of the sum of elementary
charges $Q_f$. The connected part, on the other hand is proportional to
the sum of the square of the elementary charges of the quark flavor $f$, 
$C_{em}=\sum_f Q_f^2$. 
In the following we will limit our discussion to the connected part 
of the vector current, as it is generally done in the analysis of thermal
observables deduced from the electromagnetic current.
 
The current-current correlation functions can be represented in terms of
an integral over spectral functions, $\rho_{\mu\nu}(\omega,\vec{p},T)$,
which are odd functions of the 4-momentum, {\it i.e.}
$\rho_{\mu\nu} (\omega,\vec{p},T) = - \rho_{\mu\nu} (-\omega, -\vec{p},T)$.
We denote by $\rho_{ii}$ the sum over the three
space-space components of the spectral function and also introduce the
vector spectral function $\rho_V \equiv \rho_{00}+\rho_{ii}$. With this 
we obtain the corresponding correlation functions,
\begin{equation}
G_{H}(\tau,\vec{p},T) =  \int_{0}^{\infty} \frac{{\rm d} \omega}{2\pi}\;
\rho_{H} (\omega,\vec{p},T)\;
{{\rm cosh}(\omega (\tau - 1/2T)) \over {\rm sinh} (\omega /2T)}
\quad , \quad H=00,\ ii, \ V \ .
\label{speccora}
\end{equation}

In the following we will limit ourselves to an analysis of Euclidean time 
correlation functions at vanishing three-momentum. Therefore we
will drop the second argument of the correlation and spectral
functions and also suppress the argument for the explicit temperature 
dependence, e.g. $G_{H}(\tau T) \equiv G_{H}(\tau, \vec{0},T)$.
Occasionally we also find it convenient to express dimensionful quantities in
appropriate units of temperature. For this purpose we will introduce 
dimensionless variables, e.g.
$\widetilde{\omega}=\omega/T$, $\widetilde{\tau}=\tau T$ and 
$\widetilde{G}_H = G_H/T^3$.

\subsection{Spectral functions}

The integral over the $0^{th}$ component of the vector current is the net number
of quarks ($q$-$\bar{q}$)  in a given flavor channel, 
$n_0(\tau)= \int {\rm d}^3x J_0(\tau, \vec{x})$. As the net number of 
quarks is conserved, it also does not depend on  Euclidean time $\tau$, 
{\it i.e.} $n_0(\tau) =n_0(0)$. Consequently the 
corresponding correlation function, $G_{00}(\tau T)$, is constant
in Euclidean time and its spectral representation is
simply given by a $\delta$-function,
\begin{equation}
\rho_{00}(\omega) = - 2\pi \chi_q \omega \delta (\omega) \; ,
\label{rho_00}
\end{equation}
where $\chi_q$ is the quark number susceptibility\footnote{As discussed
above we only consider the connected part of the vector correlation
functions. Within this approximation $\chi_q$ is actually not the quark number 
susceptibility but the isospin susceptibility \cite{Allton6}.}
\begin{equation}
\chi_q = - \frac{1}{T}\int {\rm d}^3x\ 
\langle J_0(\tau, \vec{x}) J_0^{\dagger}(0,\vec{0}) \rangle \; .
\label{suscept}
\end{equation}
The time-time component of the 
correlation function thus is proportional to the quark number 
susceptibility, $G_{00}(\tau T) = - \chi_q T$. 
The vector correlation functions $G_{ii}$ and $G_{V}$ therefore differ
only by a constant,
\begin{equation}
G_{ii} (\tau T) = \chi_q T + G_V (\tau T) \; .
\label{relation}
\end{equation}

The spatial components of the vector spectral function, 
$\rho_{ii}(\omega)$, increase quadratically for large values of 
$\omega$. In the free field limit one obtains for massless quarks,
\begin{eqnarray}
\rho_{00}^{\rm free} (\omega) &=&  - 2\pi T^2 \omega \delta (\omega ) \; , 
\nonumber \\
\rho_{ii}^{\rm free} (\omega) &=& 2\pi T^2 \omega \delta (\omega ) + 
{3 \over 2 \pi} \; \omega^2  \;\tanh (\omega/4T) \; .
\label{spectral_v0i}
\end{eqnarray}
Note that in this limit the contributions of $\delta$-functions in
$\rho_{00}$ and $\rho_{ii}$  cancel in the vector spectral function 
$\rho_V(\omega) \equiv \rho_{00}(\omega) +\rho_{ii} (\omega)$. 
This is no longer the case in an interacting theory at finite temperature.

At high temperature and for large energies corrections to the free field 
behavior can be calculated perturbatively; the vector spectral 
function can be deduced from the calculation of one loop corrections
to the leading order results for the thermal dilepton rate \cite{Aurenche}.
This yields,  
\begin{equation}
\rho_{ii}(\omega) \simeq 
 {3 \over 2 \pi} \left( 1 + \frac{\alpha_s}{\pi}  \right)
\; \omega^2  \;\tanh (\omega/4T) \quad , \quad
\omega /T \gg 1 \; .
\label{spectral_ii}
\end{equation}
At energies $\omega \lsim T$ perturbative calculations as well as the 
resummation of certain subsets of diagrams (hard thermal loop (HTL) resummation
\cite{Braaten}) become complicated as several scales of order $g^n T$ start
becoming important. In fact, the straightforward HTL-resummation
\cite{Braaten} is known to lead to a infrared divergent spectral function, 
$\rho_{\rm HTL} (\omega) \sim 1/\omega$, which also leads to a divergent vector
correlation function \cite{Thoma}.
 
The limit $\omega \rightarrow 0$ is sensitive to transport properties in the
thermal medium and the spectral functions need to be linear in $\omega$ in
order to give rise to a non-vanishing, finite transport coefficient; in this
limit the spatial components of the vector spectral function yield the 
electrical conductivity
\begin{equation}
\frac{\sigma}{T} = \frac{C_{em}}{6} \lim_{\omega \rightarrow 0} 
\frac{\rho_{ii}(\omega)}{\omega T} \; .
\label{conduct}
\end{equation}

In the free field, infinite temperature limit also the spatial part of the 
spectral function contains a $\delta$-function at the origin. 
Different from the time-time component, where the $\delta$-function
is protected by current conservation, this $\delta$-function is smeared
out at finite temperature and the low energy part 
of $\rho_{ii}$ is expected to be well described by a Breit-Wigner peak 
\cite{Boon,Forster,Aarts02,Teaney06,Teaney10},
\begin{equation}
\rho_{ii}^{BW}(\omega) = \chi_q c_{BW}
\frac{\omega \Gamma}{\omega^2 + (\Gamma/2)^2} 
\; ,
\label{BWpeak}
\end{equation}
which yields $\sigma(T)/C_{em} = 2 \chi_q c_{BW}/(3 \Gamma)$. In the infinite
temperature limit the width of the Breit-Wigner peak vanishes; at the same time 
$c_{BW}\rightarrow 1$, $\chi_q \rightarrow T^2$ and consequently the electrical 
conductivity is infinite in the non-interacting case.

In the high temperature regime the time-time component of the spectral 
function receives perturbative corrections, reflecting the perturbative
corrections to the quark number susceptibility. To leading order this gives
\begin{equation}
\rho_{00} = -2\pi T^2\omega \delta (\omega) \left( 1 - 
\frac{1}{2\pi^2} g^2(T) \right) \; .
\label{pert00}
\end{equation}

\subsection{Thermal dilepton and photon rates}
The vector spectral function 
is directly related to the thermal production rate of dilepton pairs
with squared invariant mass $\omega^2 - \vec p^2$, 
\begin{equation}
{{\rm d} N_{l^+l^-} \over {\rm d}\omega {\rm d}^3p} =
C_{em}{\alpha^2_{em}  \over 6 \pi^3} {\rho_V(\omega,\vec{p},T) 
\over (\omega^2-\vec{p}^2) ({\rm e}^{\omega/T} - 1)}
\quad ,
\label{rate}
\end{equation} 
where 
$\alpha_{em}$ is the electromagnetic fine structure constant.

The vector spectral function at light-like 4-momentum 
yields the photon emission rate of a thermal medium,
\begin{equation}
\omega \frac{{\rm d} R_\gamma}{{\rm d}^3p} =C_{em} \frac{\alpha_{em}}{4\pi^2} 
\frac{\rho_{V}(\omega =|\vec{p}|, T)}{{\rm e}^{\omega/T} -1} \ .
\label{photon}
\end{equation}
The emission rate of soft photons, thus can be related to the electrical
conductivity,
\begin{equation}
\lim_{\omega \rightarrow 0} \omega \frac{{\rm d} R_\gamma}{{\rm d}^3p} =
\frac{3}{2\pi^2} \sigma(T) T \alpha_{em} \ .
\label{softphoton}
\end{equation}

\section{Moments of the vector spectral function}

In addition to the vector correlation function itself we will calculate 
its curvature at the largest Euclidean time separation accessible
at non-zero temperature, {\it i.e.} at $\tau T =1/2$. The
curvature is the second 
{\it thermal moment} of the spectral functions at vanishing 
momentum,
\begin{equation}
G_H^{(n)} = \frac{1}{n!} \left. \frac{{\rm d}^n G_H(\tau T)}{{\rm d} (\tau T)^n}
\right|_{\tau T =1/2}  =\frac{1}{n!} \int_{0}^{\infty} \frac{{\rm d} \omega}{2\pi}
\left( \frac{\omega}{T}\right)^n\ { \rho_H (\omega)
\over {\rm sinh} (\omega /2T)}~~, \;\; H=ii,\ V\ ,
\label{moments}
\end{equation}
where $n$ is chosen to be even as all odd moments vanish.
These thermal moments give the Taylor expansion coefficients for the 
correlation function expanded around the mid-point of the Euclidean time 
interval, {\it i.e.} around $\tau T=1/2$,
\begin{equation}
G_H(\tau T) = \sum_{n=0}^{\infty} G_H^{(2n)} \left( \frac{1}{2} - \tau T \right)^{2n} \ .
\label{taylor}
\end{equation}
It is obvious from Eq.~\ref{moments} that in a calculation of
the n-th moment the spectral function is weighted by a term, which has a 
pronounced maximum at $\tanh (\omega /2T)=\omega /2nT$. 
The n-th order moment thus is most sensitive to the spectral function 
at $\omega/T\simeq 2n$ and  we may expect that
higher order moments will come close to results obtained with perturbative 
spectral functions.

In the infinite temperature, free field limit the integral in Eq.~\ref{speccora} 
can be evaluated analytically and one obtains for massless quarks 
\cite{Flo94},
\begin{eqnarray}
G_{V}^{free} (\tau T) &=& 
6  T^3 \left( \pi \; ( 1-2\tau T)
{1+\cos^2(2\pi \tau T ) \over \sin^3 (2\pi \tau T )} 
+ 2 \; {\cos(2\pi \tau T ) \over \sin^2 (2\pi \tau T )} \right) \quad , \nonumber \\
G_{ii}^{free} (\tau T) &=& T^3 + G_{V}^{free} (\tau T) \; .
\label{cor_free}
\end{eqnarray}
From this we easily obtain the moments of the free spectral functions. The first
three non-vanishing moments are given by
\begin{equation}
G_V^{(0),free} = \frac{2}{3} G_{ii}^{(0),free} = 2 T^3 \; ,\;\; 
G_H^{(2),free} = \frac{28 \pi^2}{5} T^3
\; ,\;\; G_H^{(4),free} = \frac{124 \pi^4}{21} T^3\; .
\label{free_moments}
\end{equation} 

\noindent
We note that $G_{ii}^{(n)}=G_V^{(n)}$ for all $n > 0$ 
not only at an infinite but at all values of the 
temperature as the correlators $G_{ii}$ and $G_V$ only differ by a constant 
(Eq.~{\ref{relation}}). 
However, while in the free field limit 
$G_{ii}(\tau T)$ contains a constant contribution which drops
out in the calculation of 
higher moments ($n>0$),
this is no longer the case at finite values of the temperature:
all thermal moments, $G_H^{(2n)}$ with $n\ge 0$, will be sensitive to the 
smeared  $\delta$-function contributing to $\rho_{ii}(\omega)$ although, 
as indicated above, we expect this contribution to become more and more 
suppressed in higher order moments.

In the following we will analyze the ratio of $G_H(\tau T)$ and the free
field correlator $G_H^{free}(\tau T)$,
\begin{eqnarray}
\frac{G_H(\tau T)}{G_H^{free}(\tau T)} 
&=& \frac{G_H^{(0)}}{G_H^{(0),free}}
\left( 1+ \left( R^{(2,0)}_H -R^{(2,0)}_{H,free} \right) 
\left(\frac{1}{2} - \tau T \right)^{2} + ....\right) \; ,
\label{series}
\end{eqnarray}
as well as the ratio of mid-point subtracted correlation functions
\begin{eqnarray}
\Delta_V(\tau T) &\equiv& 
\frac{G_V(\tau T)-G_V^{(0)}}
{G_V^{free}(\tau T)-G_V^{(0),free}}  \nonumber \\
&=& 
\frac{G^{(2)}_V}{G^{(2),free}_{V}}
\left( 1+ \left( R^{(4,2)}_V -R^{(4,2)}_{V,free} \right)
\left(\frac{1}{2} - \tau T \right)^{2} + ....\right) \; .
\label{mid-point}
\end{eqnarray}
Here we used the notation $R_H^{(n,m)} \equiv G_H^{(n)}/G_H^{(m)}$.
Note that the curvature of these ratios at the mid-point determines the 
deviation of ratios of thermal moments from the
corresponding free field values. While the ratios of 
correlation functions differ in the $H=ii$ and $H=V$ channels due to 
the additional constant contributing to $G_{V}(\tau T)$, this constant 
drops out in the subtracted correlation function, {\it i.e.}, 
$\Delta_V(\tau T)\equiv \Delta_{ii}(\tau T)$ and $G_V^{(n)}= G_{ii}^{(n)}$
for $n>0$. 

\section{Computational details and numerical results}

All our numerical results are obtained from an analysis of quenched QCD
gauge field configurations generated with the standard SU(3) single plaquette
Wilson gauge action \cite{Wilson}.  On these gauge field configurations
correlation functions have been calculated using a clover action with 
non-perturbatively determined clover coefficients $c_{SW}$
and a hopping parameter $\kappa$  chosen close to its critical value 
\cite{Luscher1,Luscher}. 

\begin{table}[t]
\begin{center}
\vspace{0.3cm}
\begin{tabular}{|l|c|c|c|c|c|}
\hline
 $N_\tau$ & $\beta$ & $T/T_c$ & $c_{SW}$ & $\kappa$ & $Z_V$ \\
\hline
16 &  6.872 & 1.46 & 1.4125 & 0.13495 & 0.829 \\
\hline
24 (I) &  7.192 & 1.42 & 1.3673 & 0.13431 & 0.842  \\
24 (II) &    &   &   & 0.13440 &   \\
\hline
32 &  7.457 & 1.45 & 1.3389 & 0.13390 & 0.851 \\
\hline
48 &  7.793 & 1.43 & 1.3104 & 0.13340 & 0.861 \\
\hline
\end{tabular}
\end{center}
\caption{Simulation parameters for the generation of gauge field 
configurations on lattices of size $N_\sigma^3\times N_\tau$.
}
\label{tab:parameter}
\end{table}

The gauge couplings used for calculations at four
different values of the cut-off ($aT=1/N_\tau$) have been selected such that
the temperature stays approximately constant as the cut-off is varied;
$T\simeq 1.45 T_c$. 
For this we used an ansatz originally suggested in Ref.~\cite{Allton}.
It is known to give the variation of the lattice cut-off as function of the 
gauge coupling to better than 1\% in the interval $[5.6,6.5]$ \cite{Edwards}.
We added to this analysis new results 
for the critical coupling $\beta_c(N_\tau)$ and the square 
root of the string tension $\sqrt{\sigma}$ \cite{Lucini} 
and extrapolated the fit results to the regime of 
couplings relevant for our analysis, {\it i.e.} $\beta \in [6.8,7.8]$.

\begin{table}[t]
\begin{center}
\vspace{0.3cm}
\begin{tabular}{|c|c|c|c|c|c|c|}
\hline
$\beta$ & $\kappa$ & $m_{AWI}/T[\mu=1/a]$ & $m_{\overline{{\mathrm MS}}}/T[\mu=2GeV]$ & $m_{\overline{{\mathrm MS}}}[\mu=2GeV]\ [{\rm MeV}]$ \\
\hline
6.872 &  0.13495 & 0.0182(4) & 0.0243(1) & 9.5\\
7.192 &  0.13440 & 0.0182(4) & 0.0237(1) & 9.3\\
7.192 &  0.13431 & 0.0816(2) & 0.1062(2) & 41.6\\
7.457 &  0.13390 & 0.0790(3) & 0.0989(4) & 38.7\\
7.793 &  0.13340 & 0.0964(2) & 0.1117(2) & 43.7\\
\hline
\end{tabular}
\end{center}
\caption{
AWI and $\overline{{\mathrm MS}}$ quark masses in units of the temperature.
For orientation we show in the last column estimates for the quark masses in
units of MeV. These values have been obtained using for the phase transition
temperature in the SU(3) gauge theory $T_c\simeq 270$~MeV.}
\label{tab:quark_masses}
\end{table}

\begin{table}[t]
\begin{center}
\vspace{0.3cm}
\begin{tabular}{|c|c|c|c|}
\hline
 $N_\tau$ & $N_\sigma$ & $\beta$ & \# conf \\
\hline
16 &  32 &  6.872 &  251 \\
~  &  48 &  6.872 &  229 \\
~  &  64 &  6.872 &  191 \\
~  &  128 &  6.872 & 191 \\
\hline
24 (I) &  128 &  7.192 & 340 \\
24 (II)&  128 &  7.192 & 156 \\
\hline
32 &  128 &  7.457 & 255 \\
\hline
48 &  128 &  7.793 & 451\\
\hline
\end{tabular}
\end{center}
\caption{Number of configurations analyzed on
lattices of size $N_\sigma^3\times N_\tau$. All configurations
are separated by 500 Monte Carlo iterations.
}
\label{tab:statistics}
\end{table}

By tuning the Wilson hopping parameter $\kappa$ in the fermion action we have
chosen quark masses that are approximately constant for our four values 
of the cut-off.
We have estimated the quark masses using the axial Ward identity (AWI)
to calculate the so called AWI current quark mass, $m_{AWI}$, for the
different values of the cut-off \cite{AWI}. Here we used a non-perturbatively 
improved axial-vector current with coefficient $c_A$ from 
\cite{Luscher1,Guagnelli:1997db}. 

To compare the quark masses at a common scale we first converted
to renormalization group invariant quark masses \cite{Capitani:1998mq}
using non-perturbative coefficients \cite{deDivitiis:1997ka,Guagnelli:2000jw} 
and then rescaled in the $\overline{{\mathrm MS}}$ scheme to the common scale
$\mu=2$~GeV using the four-loop perturbative running from 
\cite{Chetyrkin:2000yt}.
In both cases errors are calculated from a jackknife analysis on
$m_{AWI}$ and do not include systematic errors for the conversion to
the $\overline{{\mathrm MS}}$ scheme.

All basic parameters that enter our simulations are summarized in 
Table~\ref{tab:parameter}.
The values for the quark masses used in our calculations are quoted  in 
Table~\ref{tab:quark_masses}. Our central results are based on calculations
performed with renormalized quark masses $m_{\overline{{\mathrm MS}}}/T \simeq 0.1$.
Through calculations performed with a factor 4 smaller quark mass we checked
on lattices with temporal extent $N_\tau =24$ that results
on the large distance behavior of correlation functions agree within errors
(see Fig.~\ref{fig:GV_volume}). 
This smaller quark mass has also been used for the calculations on our 
coarsest lattice corresponding to a temporal lattice size $N_\tau=16$, where
we analyzed the sensitivity of our results to finite volume effects.
We thus expect that quark mass effects play no role in all results 
presented here.

We have generated the gauge field configurations 
by using an over-relaxed heat bath algorithm.
Configurations have been stored after every 500 sweeps. 
We have checked at several large temporal separations that
the correlation functions calculated on these
configurations are statistically independent.
We further have measured the plaquette every sweep to obtain
an estimate for the integrated autocorrelation time of order 1.
Note that at temperatures above deconfinement topologically
non-trivial configurations 
die out quickly. This is reflected in the observation that the inspection
of the scalar and pseudo-scalar correlator revealed less than 
3 \% so-called exceptional
configurations\footnote{Note that zero modes do not couple into the 
vector and axial vector correlation functions.}
which had an autocorrelation time of 3000 sweeps at most.
The statistics accumulated on different size lattices is collected
in Table~\ref{tab:statistics}.

For the computation of the quark propagators we used a plain
conjugate gradient inverter. The algorithm has a stable convergence
behavior, however, we found it appropriate to set the convergence criterion to 
$10^{-23}$ for the squared norm of the residue. This value was selected
by monitoring the nearly exponential decay of spatial correlation functions
which range over many orders of magnitude due to the large spatial
extent of the lattices and screening masses of the order of $2 \pi T$.

For the calculation of the time-time ($G_{00}$) and space-space ($G_{ii}$) 
components of the vector correlation function we use the local
vector current $J_\mu (\tau,\vec{x})$ introduced in Eq.~{\ref{V_current}}. 
Unlike point-split vector currents, this local lattice current 
is not conserved at non-zero lattice spacing.
As a consequence  
the time-time component $G_{00}(\tau T)$ may not strictly be $\tau$-independent 
at finite values of the cut-off {\cite{Aarts05}}.
Furthermore, the current needs to be renormalized multiplicatively.
Wherever required we therefore use the renormalized
vector current,
\begin{equation}
J^{lat}_\mu (\tau,\vec{x}) = (2 \kappa Z_V) 
\bar{q}(\tau, \vec{x})\gamma_\mu q (\tau, \vec{x}) \ ,
\label{V_renorm}
\end{equation}
with 
$Z_V$ denoting
the non-perturbatively determined renormalization constant {\cite{Luscher}}
also given in Table~{\ref{tab:parameter}}.

However, we note that many results presented in the following are given  
in terms of ratios of correlation functions as well as mid-point
subtracted correlation functions divided by the quark number
susceptibility, $\widetilde{\chi}_q\equiv\chi_q/T^2$. 
These ratios are independent of any multiplicative renormalization.

It is obvious from Eq.~\ref{series} that the correlators $G_V(\tau T)$ and
$G_{ii}(\tau T)$ are identical up to a constant,
$G_{00}(\tau T)\equiv \chi_q T^2$. It thus suffices to analyze one of them.
In the following we will mainly concentrate on a discussion of
$G_V(\tau T)$. This is because a subtle cancellation occurs between the
constant contribution $\chi_q T^2$ to $G_V(\tau T)$ and the
Breit-Wigner term in the spatial part ($G_{ii}(\tau T)$). In the free field
limit this cancellation is exact. It is the incomplete cancellation of these
two contributions to $G_V(\tau T)$ which gives rise to the weak dependence of
$G_V/G_V^{free}$ on Euclidean time and contains information on transport
coefficients. Of course, the constant contribution $\chi_q T^2$ is trivial
when it comes to an analysis of the spectral representation of vector
correlation functions. For the determination of the spectral function
$\rho_{ii}(\omega)$ the correlator $G_{ii}(\tau T)$ thus is much more
relevant. For this reason we will also show some results on the spatial
correlator.
 
\subsection{Correlation functions of the vector current}

The vector current correlation function was calculated previously, 
using the same non-perturbatively improved clover action, on lattices of 
size $64^3\times 16$ \cite{our_dilepton}. A MEM analysis was used 
to determine thermal dilepton rates which suggested a suppression of
the spectral weight relative to the free spectral function at small
energies. 
Subsequently vector correlation functions at high temperature were
analyzed using staggered fermion formulations \cite{Gupta,Aarts07} 
and a modified kernel was used in the MEM analysis \cite{Aarts07}. This 
modification of the MEM algorithm increased the sensitivity of MEM to the 
low energy structure of spectral functions and resolved part of the problems 
observed in  \cite{our_dilepton}. We will show below that it is also
crucial in a MEM analysis to choose default models which allow for a 
linear slope of the spectral functions at small energies. 

Calculations of vector spectral functions using staggered fermions are
more involved, as two spectral functions, corresponding to different parity
channels, need to be determined simultaneously.
These studies were performed using lattices with temporal extent
up to $N_\tau =14$ \cite{Gupta} and $N_\tau=24$ \cite{Aarts07}, respectively.
They were performed with unrenormalized currents and primarily
aimed at a determination of the (unrenormalized) electrical conductivity. 
They led, however, to quite different results, 
$\sigma/T \simeq 7 C_{em}$ \cite{Gupta} and 
$\sigma/T \simeq (0.4\pm 0.1) C_{em}$
\cite{Aarts07}. 

Here we improve over these studies with staggered fermions as well as  
the analysis 
of the thermal dilepton rate performed with improved Wilson fermions 
\cite{our_dilepton} by using the latter approach and doubling the
spatial lattice extent to $N_\sigma=128$ as well as increasing the 
temporal lattice extent by up to a factor three. 
We have calculated the vector correlation function on lattices 
of size
$N_\sigma^3\times N_\tau$, with $32 \le N_\sigma \le 128$ and $N_\tau = 16,\
24,\ 32$ and $48$. 
For $N_\tau=16$ we have calculated $G_H(\tau T)$ on lattices
with spatial extent $N_\sigma = 32,\ 64,\ 96$ and $128$. For 
$N_\tau =24$ we checked that the quark masses used in our calculations
are indeed small enough on the scale of the temperature to be ignored
in the analysis of our correlation functions.
On the largest spatial lattice, $N_\sigma =128$, we performed calculations
for four different values of the lattice cut-off by choosing $N_\tau= 16,\
24,\ 32$ and $48$ and at the same time changing the value of the gauge coupling
$\beta$ such that the temperature is kept constant, $T\simeq 1.45 T_c$.

The large range of spatial lattice sizes used in this calculation, 
$2\le N_\sigma/N_\tau \le 8$, allows to quantify finite volume effects 
at fixed values of the lattice cut-off,  $aT=1/N_\tau$. The large temporal 
lattice allows to reduce the lattice spacing at $T\simeq 1.45 T_c$ to about 
$0.01$~fm. Its variation by a factor three gives us good 
control over lattice cut-off effects. As will become clear in the following,
finite volume effects are well under control and the variation of the lattice 
cut-off by a factor three now allows for the extrapolation 
to the continuum limit. This removes systematic errors that were present in
earlier calculations performed with the same discretization scheme used here
\cite{our_dilepton}.

\begin{figure}
\begin{center}
\epsfig{file=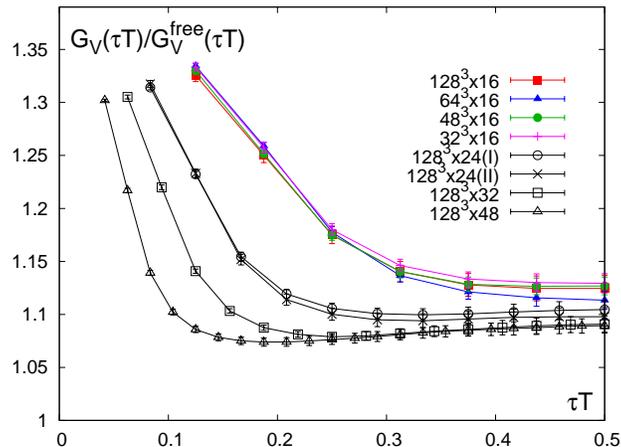,width=90mm}
\end{center}
\caption{The vector correlation function, $G_V(\tau T)$, 
calculated on lattices of size $N_\sigma^3\times N_\tau$
at $T\simeq 1.45T_c$. The correlation functions are normalized
to the free vector correlation function $G_V^{free}(\tau T)$ in the continuum.
Shown are data for $\tau T >1/N_\tau$ only. Label I and II refer to  
data sets generated with two different values of the quark mass 
(see Table~\ref{tab:parameter}). 
}
\label{fig:GV_volume}
\end{figure}

\begin{figure}[t]
\begin{center}
\epsfig{file=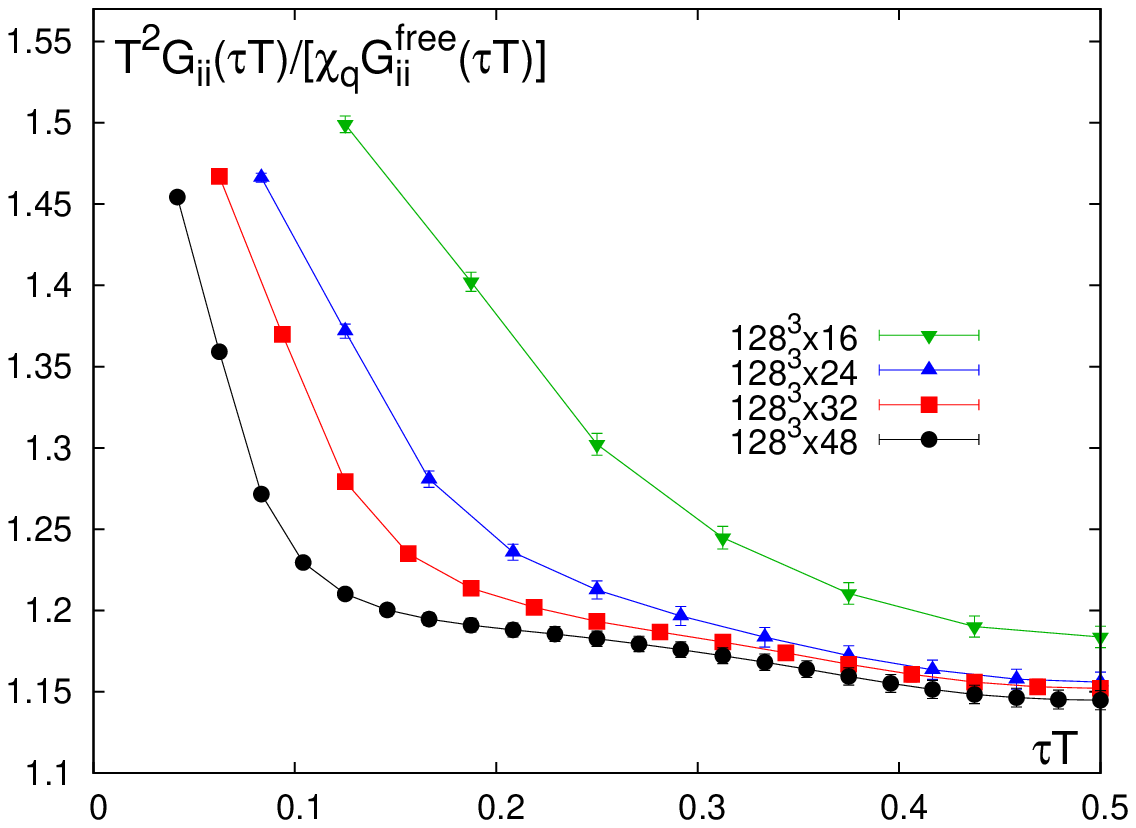,width=78mm}\hspace*{-0.8cm}
\epsfig{file=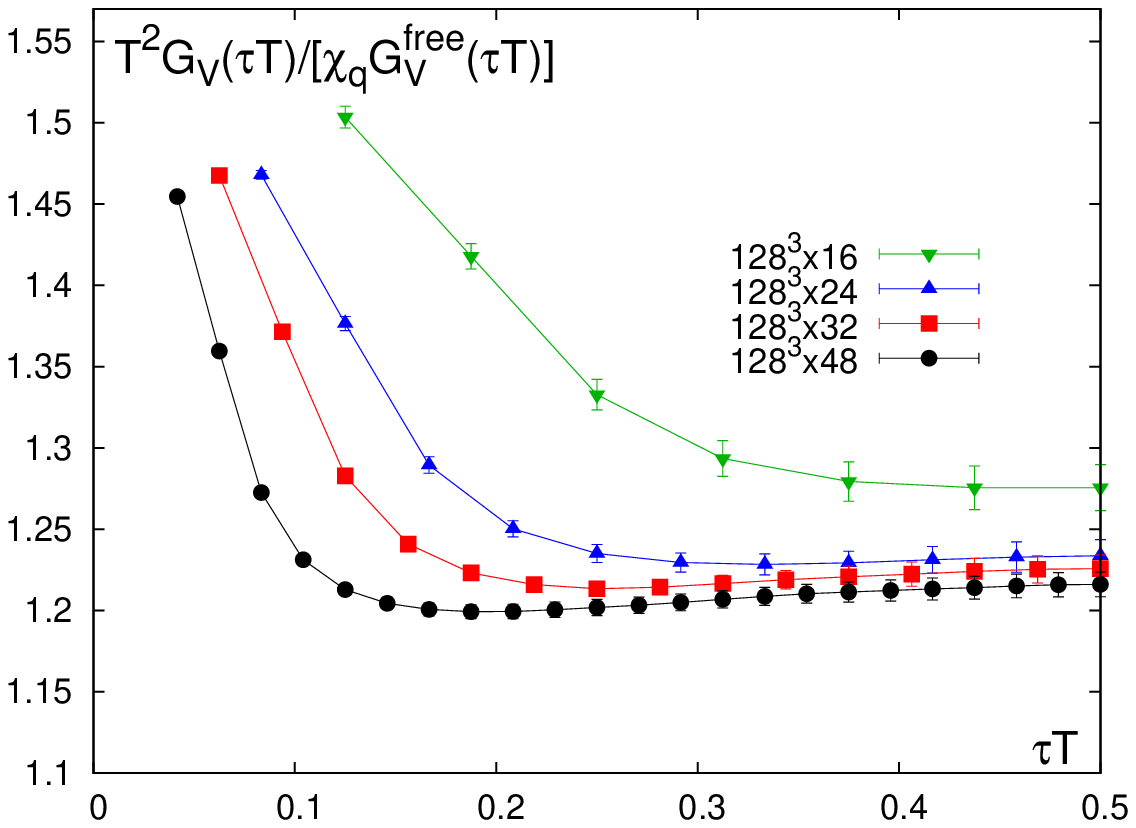,width=78mm}

\epsfig{file=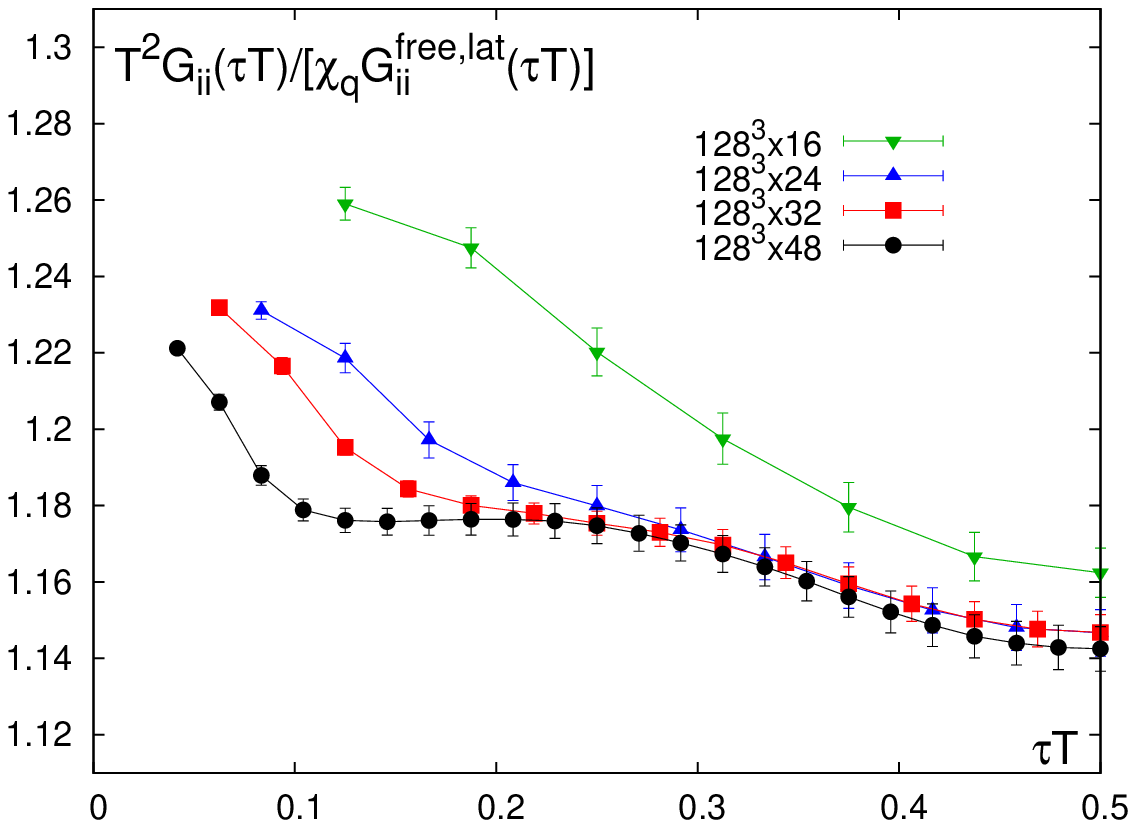,width=78mm}\hspace*{-0.8cm}
\epsfig{file=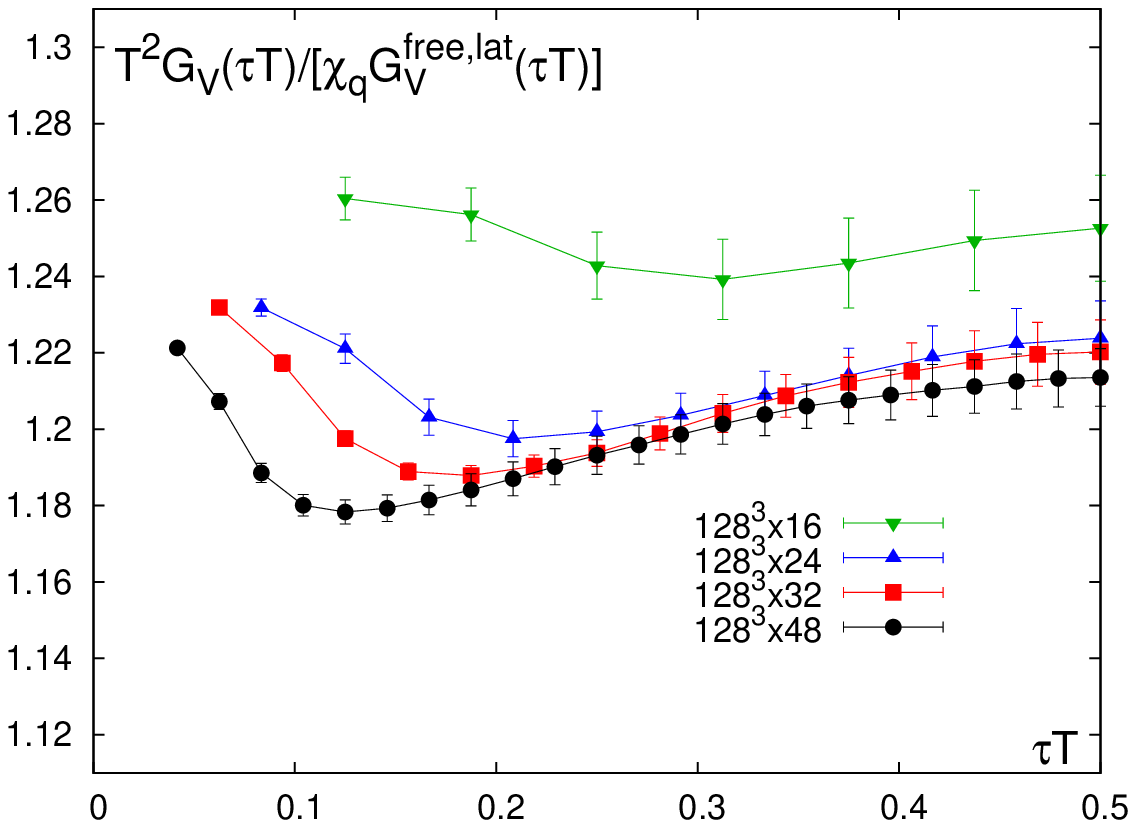,width=78mm}
\end{center}
\caption{The space-space component of the vector correlation function, 
$G_{ii}(\tau T)/T^3$, calculated at $T\simeq 1.45T_c$ on lattices of size
$128^3\times N_\tau$ versus Euclidean time $\tau T$ 
(left) and the total vector correlator $G_{V}(\tau T)/T^3$ (right).
Both correlators have been normalized by the continuum version of the 
corresponding free vector 
correlation function defined in Eq.~\protect\ref{cor_free} (upper part)
and the discretized  version of the free vector correlation function
calculated on infinite spatial volumes with fixed $N_\tau$ (lower part),
respectively.
}
\label{fig:GV}
\end{figure}

As the dependence of $G_H(\tau T)$ on Euclidean time $\tau$ is very similar 
to that of the free vector correlation function, we eliminate the dominant 
exponential variation of $G_H(\tau T)$ with $\tau T$ and show in 
Fig.~\ref{fig:GV_volume} the ratio $G_V(\tau T)/G_V^{free}(\tau T)$ 
introduced in Eq.~\ref{series} for all parameter sets analyzed by us.
We note that finite volume effects increase somewhat with 
$\tau T$, but remain small even for the largest 
Euclidean time separation $\tau T=1/2$. 
Cut-off effects show the opposite trend; they increase with 
decreasing $\tau T$. In fact,
they strongly influence the behavior of $G_V(\tau T)$
at short distances for the first 6 to 8 Euclidean time units, 
$1/N_\tau \le \tau T \ \lsim\  8/N_\tau$. It also is obvious from 
Fig.~\ref{fig:GV_volume}
that deviations from the free field correlation function change qualitatively
when the lattice spacing is reduced. 
For $N_\tau \ge 24$ the ratio $G_V(\tau T)/G_V^{free}(\tau T)$
increases close to $\tau T=1/2$ 
indicating that the ratio of thermal moments, 
$R^{(2,0)}_V$, is smaller than the corresponding free field value, 
$R^{(2,0)}_V< R^{(2,0)}_{V,free}$. The spatial
part alone, $G_{ii}(\tau T)/G_{ii}^{free}(\tau T)$, on the other hand drops
when approaching $\tau T=1/2$ (see Fig.~\protect\ref{fig:GV}). 
This suggests $R^{(2,0)}_{ii} > R^{(2,0)}_{ii,free}$. 

\begin{figure}
\epsfig{file=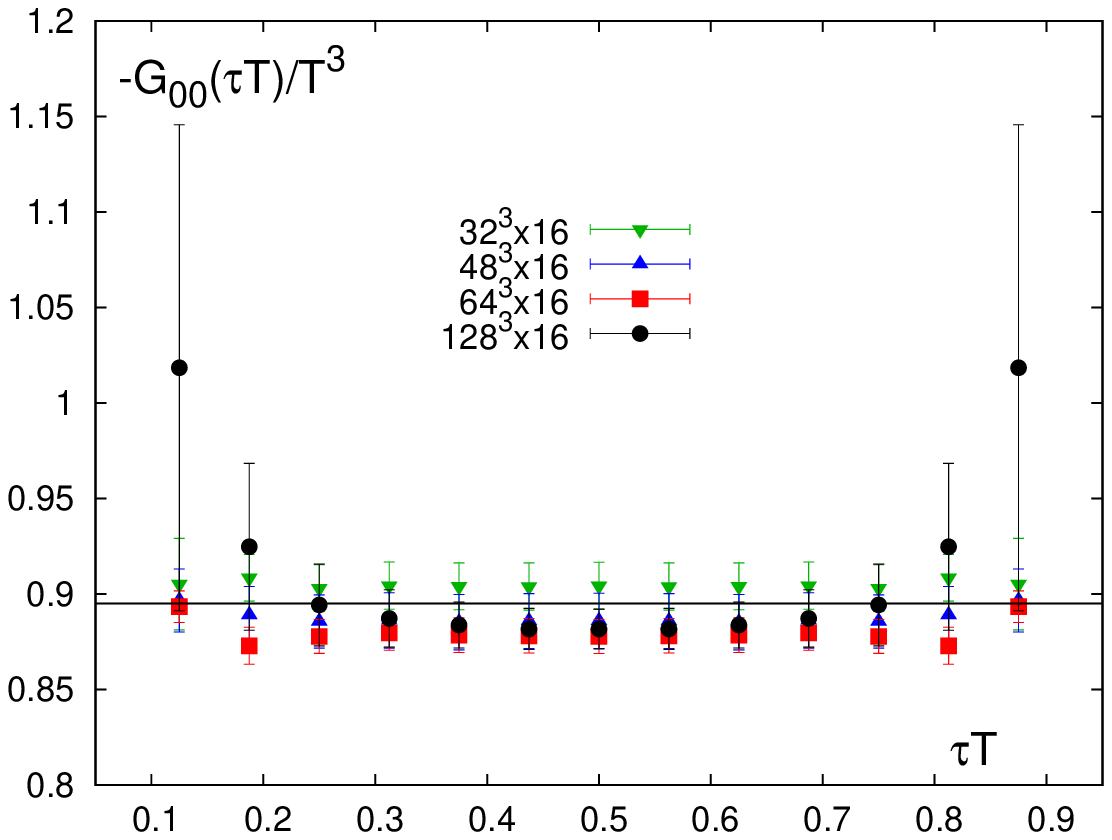,width=78mm}\hspace*{-0.8cm}
\epsfig{file=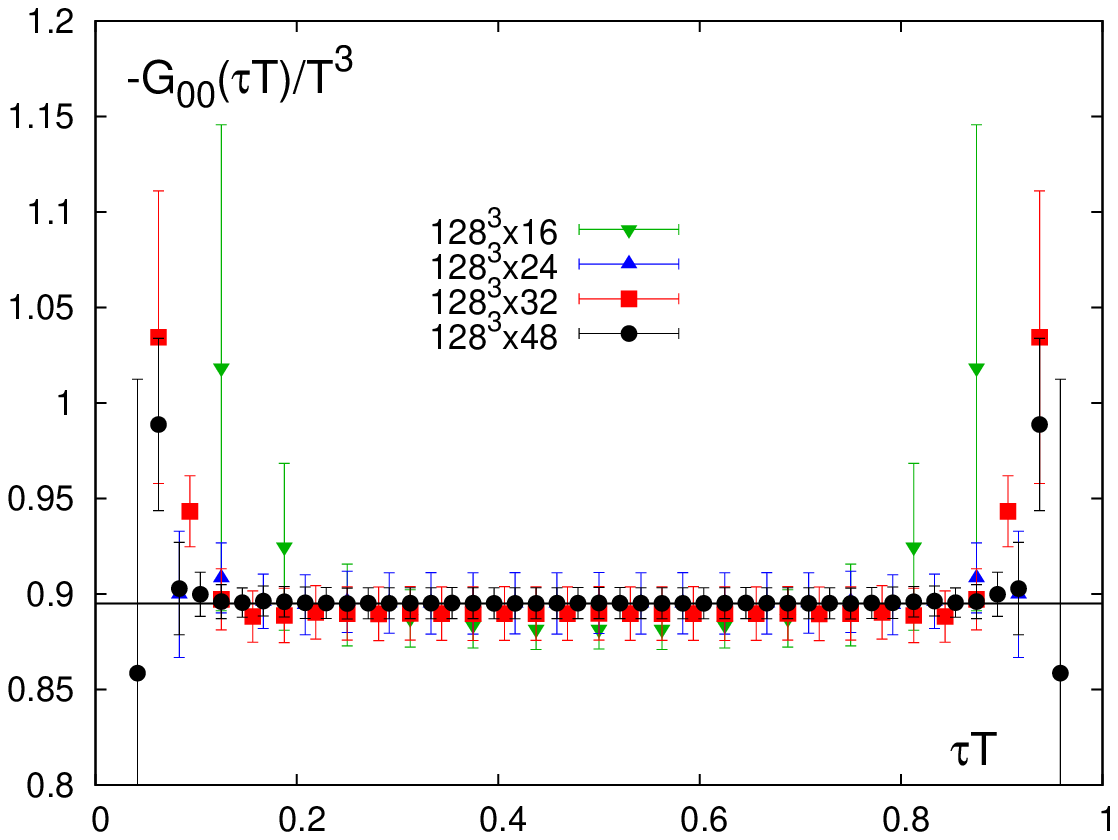,width=78mm}
\caption{The time-time component of the vector correlation function, 
$G_{00}(\tau T)/T^3$, calculated at $T\simeq 1.45T_c$ versus Euclidean time $\tau T$. 
The left hand part of the
figure shows the volume dependence of $G_{00}(\tau T)/T^3$ for $N_\tau=16$ and 
$32 \le N_\sigma\le 128$.
The right hand figure shows the cut-off dependence of $G_{00}(\tau,T)/T^3$ 
for $N_\sigma =128$ and $16\le N_\tau \le 48$. In both figures the horizontal
line shows a fit to the data obtained on the largest lattice, $128^3\times 48$.
}
\label{fig:G00}
\end{figure}

Also in calculations of the  time-time component of the vector correlation 
function we find that finite volume, as well as cut-off effects, are small. 
In Fig.~\ref{fig:G00} we show results for $-G_{00}(\tau T)/T^3$
for all lattice sizes analyzed by us.
The left hand part of Fig.~\ref{fig:G00} shows results obtained on lattices with
different spatial extent at fixed lattice cut-off. 
Except for the smallest aspect ratio, $N_\sigma/N_\tau=2$, the results agree 
within statistical errors of about 1\%.  The right hand part shows
results obtained on our largest spatial lattice, 
$N_\sigma =128$, at four different values of the lattice cut-off. 
This shows that also cut-off effects are small in the calculation of 
the time-time component of the vector correlation function, {\it i.e.} the
quark number susceptibility. Apparently the fact that we use a non-conserved 
local current does not significantly alter the spectral properties of the 
time-time correlator. Except for short distances it is  to a good degree 
$\tau$-independent. We summarize results for $\chi_q/T^2 \equiv
-G_{00}(\tau T)/T^3$ calculated on the $128^3\times N_\tau$ lattices in 
Table~\ref{tab:G00}.

\begin{table}[t]
\begin{center}
\vspace{0.3cm}
\begin{tabular}{|c|c|c|c|c|c|}
\hline
$N_\tau$ &  
$16$ & $24$ & $32$ & $48$ & $\infty$\\
\hline
$\chi_q/T^2$ & 0.882(10) & 0.895(16) & 0.890(14) & 0.895(8) & 0.897(3) \\
\hline
$G_V^{(2)}/(\widetilde{\chi}_q G_V^{(2),free})$
&1.273(4)  &1.214(2)  &1.207(1)  &1.193(1)  & 1.189(13) \\
\hline
$G_V(1/2)/(\widetilde{\chi}_q G_V^{free}(1/2))$
&1.276(14)  &1.234(10)  &1.226(10)  &1.216(8)  & 1.211(9) \\
$G_V(1/4)/(\widetilde{\chi}_q G_V^{free}(1/4))$
&1.333(9)  &1.235(6)  &1.213(4)  &1.202(5)  & 1.190(7) \\
\hline
$G_{ii}(1/2)/(\widetilde{\chi}_q G_{ii}^{free}(1/2))$
&1.184(7)  &1.156(6)  &1.152(5)  &1.145(6)  & 1.142(9) \\
$G_{ii}(1/4)/(\widetilde{\chi}_q G_{ii}^{free}(1/4))$
&1.302(7)  &1.213(6)  &1.193(3)  &1.183(5)  & 1.172(7) \\
\hline
\end{tabular}
\end{center}
\caption{Quark number susceptibility ($\chi_q/T^2$), 
the curvature of the vector correlation function at the mid-point
($G_V^{(2)}$) and several values of the vector correlation functions
expressed in units of the corresponding free field values and normalized
with the quark number susceptibility. Results are from calculations 
on lattices of size $128^3\times N_\tau$. The last column gives the
continuum extrapolated results.
The quark number susceptibility has been renormalized 
using the renormalization constants listed in Table~\ref{tab:parameter}.
In all other ratios renormalization constants drop out.
}
\label{tab:G00}
\end{table}

We note that at $T\simeq 1.45 T_c$ the quark number susceptibility, 
$\chi_q/T^2$, is about 10\% smaller  than the free field value 
$\chi_q^{free}/T^2 = 1$ which is in accordance with calculations performed 
with staggered fermions.  

In Fig.~\ref{fig:GV} we show separately results for the space-space component 
of the vector current correlation function, $G_{ii}(\tau T)/T^3$, and the combined 
vector correlator, $G_{V}(\tau T)/T^3$. In both cases we normalized the correlation 
function to the corresponding free correlator, $G_H^{free}(\tau T)/T^3$. 
Of course, ratios for $H=V$ and $H=ii$  are related through Eq.~\ref{relation} 
and provide identical information
on the vector spectral function representing these correlators.

It is apparent that the short distance part of the vector correlation functions
is strongly influenced by finite cut-off effects. 
To some extent one can eliminate cut-off effects by 
calculating the free field correlation functions on
lattices with finite temporal extent and
considering the ratio
$G_{V}/G_V^{free,lat}$ rather than the ratio obtained by normalizing with
the free vector correlator in the continuum limit. 
A comparison between the lower and upper panels of Fig.~\ref{fig:GV}
indeed shows a reduction of cut-off effects, although
the short distance part $\tau T \lsim 0.2$ still 
deviates substantially from the free field values. 
However, in the entire Euclidean
time interval, where cut-off effects are under control and a continuum
extrapolation is possible for $\tau T\in [0.2:0.5]$,
$G_V(\tau T)$ as well as $G_{ii}(\tau T)$ stay close to the
corresponding free field correlation functions. We also note 
that in the continuum limit 
the ratios shown in Fig.~\ref{fig:GV} will all
approach the same value in the limit $\tau T \rightarrow 0$ as 
all correlation functions diverge for $\tau T\rightarrow 0$ and the
fact that $G_V$ and $G_{ii}$ differ by a constant becomes unimportant.
Moreover, note that the data shown in this figure have been divided by 
the quark number susceptibility, $\chi_q/T^2$, to avoid the usage of
any renormalization constants.
Multiplying the data shown in Fig.~\ref{fig:GV} by the quark number 
susceptibility given in Table~\ref{tab:G00} gives the ratio of the
vector correlation functions and the free vector correlator. These 
values are about 10\% smaller than the data shown in 
Fig.~\ref{fig:GV}.  We therefore expect that in the continuum limit
deviations of the vector correlation function as well as its spatial 
component from the corresponding free correlator remain smaller than 
9\% for all $\tau T\ge 0.2$.

\subsection{Continuum extrapolation of the vector correlation function}

\begin{figure}[t]
\begin{center}
\epsfig{file=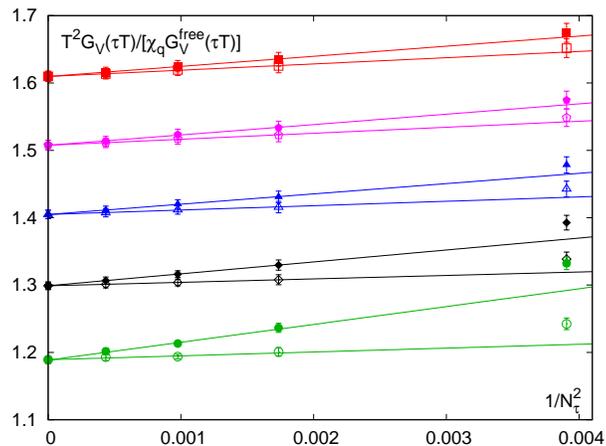,width=88mm}
\end{center}
\caption{The ratio of the vector 
correlation function, $G_{V}(\tau,T)$, normalized with the quark number 
susceptibility and the free vector correlation function calculated in the 
continuum (full symbols) and on lattices with temporal extent $N_\tau$ 
(open symbols). 
Shown are results for five values of Euclidean 
time, $\tau T = 0.1875,\ 0.25,\ 0.3125,\ 0.375$ and $0.5$ (bottom to top), 
on lattices with temporal extent $N_\tau = 16,\ 24,\ 32$ and $48$. 
For distances larger than $\tau T = 0.1875$ data have been shifted in
steps of $0.1$ for better visibility. For $\tau T = 0.1875$, and $0.3125$  
spline interpolations have been used on the $N_\tau =24$ lattice to 
estimate results at these Euclidean time separations. Note that the far most 
right data set, corresponding to $N_\sigma =16$, has not been included in the
extrapolation.
}
\label{fig:GV_extrapolation}
\end{figure}

\begin{figure}
\begin{center}
\epsfig{file=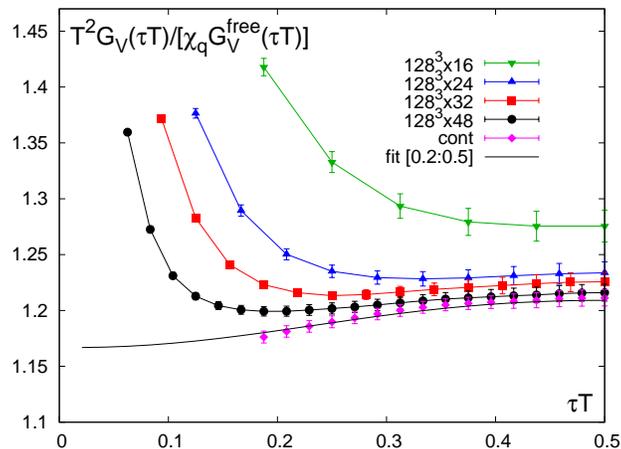,width=90mm}
\end{center}
\caption{Continuum extrapolation of the vector correlation function.
}
\label{fig:GVcont}
\end{figure}

For three values of Euclidean time, $\tau T =1/4,\ 3/8,\ 1/2$, we have 
numerical results from calculations at all four values of the lattice 
cut-off. In these cases we performed directly a continuum extrapolation of
the correlation functions using data from simulations on lattices with
temporal extent $N_\tau =24,\ 32$ and $48$. 
Specifically, we extrapolated 
the ratio of correlation functions
$G_H(\tau T)/ (\widetilde{\chi}_q G_{V}^{free}(\tau T))$
by using a quadratic ansatz in $aT=1/N_\tau$.
At other values of $\tau T$ we use spline interpolations of the data
sets on the $N_\tau =24$ and $32$ lattices to perform continuum 
extrapolations at Euclidean time distances available on the 
$N_\tau =48$ lattices, {\it i.e.} $\tau T= k/48$ for  $9\le k\ \le 24$. 
Continuum extrapolations at a few selected Euclidean time separations are 
shown in Fig.~\ref{fig:GV_extrapolation}.
We find that results obtained from the continuum
extrapolation of correlators normalized with the free lattice and continuum
correlation functions, respectively, agree within errors.
The resulting continuum
extrapolation of the vector correlation function obtained in this way is 
shown in Fig.~\ref{fig:GVcont}. Data for $\tau T=1/4$ and $1/2$ are also
summarized in Table~\ref{tab:G00}.

From Fig.~\ref{fig:GVcont} we conclude that the largest deviation 
of $G_V(\tau T)$ from 
the free vector correlation function occurs at $\tau T =1/2$. 
Taking into account the normalization with $\widetilde{\chi}_q$ 
we obtain from Table~\ref{tab:G00}
\begin{eqnarray}
\frac{G_V(1/2)}{G_V^{free}(1/2)} &=& 1.086 \pm 0.008 \; ,
\nonumber \\
\frac{G_V(1/4)}{G_V^{free}(1/4)} &=& (0.982 \pm 0.005) 
\frac{G_V(1/2)}{G_V^{free}(1/2)} \; 
\label{Gvcont}
\end{eqnarray}
where the second relation has been obtained from a jackknife analysis
of the ratio $G_V(1/4)/G_V(1/2)$.

We note that the increase of $G_V(\tau T)/G_V^{free}(\tau T)$ with $\tau T$ 
is significant. It becomes apparent only for sufficiently small lattice 
spacing, {\it i.e.} for large $N_\tau$, in particular in the normalization
with the free continuum vector correlator. The rise with $\tau T$ is a 
direct indicator that the vector spectral function in the low-$\omega$
region is different from the free case, and is part of the motivation
for the fit ans\"atze discussed in Section V.

\subsection{Curvature of the vector current correlation function}

Here we want to discuss the calculation of the second and fourth thermal
moment of the vector spectral function defined in Eq.~\ref{moments}.
From Eq.~\ref{series} we expect that the second thermal
moment will be closer to the free field value than the correlation
function at the mid-point itself.

\begin{figure}
\begin{center}
\epsfig{file=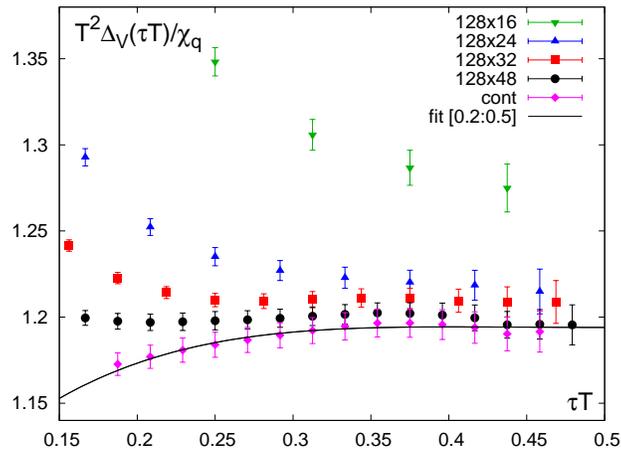,width=90mm}
\end{center}
\caption{The mid-point subtracted vector correlation function normalized
to the corresponding difference for the free vector correlation function.
Shown is the ratio as defined in Eq.~\ref{mid-point} but normalized by the
quark number susceptibility. 
The continuum extrapolation is discussed in the text.
}
\label{fig:ratios}
\end{figure}

We analyze the ratio of the mid-point subtracted correlation function,
$\Delta_V(\tau T)$, introduced in Eq.~\ref{mid-point}. Results for 
$\Delta_V(\tau T)$ calculated at the four different values of the 
lattice-cut-off are shown in Fig.~\ref{fig:ratios}.  
Again we perform 
spline interpolations of results obtained on lattices with temporal 
extent $N_\tau=24$ and $32$. These interpolated data together
with the results obtained on the $N_\tau=48$ lattices are then extrapolated 
to the continuum limit taking into account corrections of ${\cal O} ((aT)^2)$.
The extrapolated data at distances $k/48$ are shown also in 
Fig.~\ref{fig:ratios}. These extrapolated data have been fitted to a 
quartic polynomial as indicated by the Taylor expansion given in
Eq.~\ref{mid-point}. 
From this fit we obtain
\begin{eqnarray}
\frac{G_V^{(2)}}{G_V^{(2),free}} &=& 1.067\pm 0.012 \quad . \quad
\label{GV2}
\end{eqnarray}
Fits of the vector correlation function and the mid-point subtracted 
correlator are, of course, correlated. We made use of this and also calculated
the ratio of the second thermal moment and the
correlation function at the mid-point on jackknife blocks . We find
\begin{eqnarray}
\frac{G_V^{(2)}}{G_V^{(0)}} &=& (0.982\pm 0.012) 
\frac{G_V^{(2),free}}{G_V^{(0),free}}  \; , \nonumber \\
\frac{G_{ii}^{(2)}}{G_{ii}^{(0)}} &=& (1.043\pm 0.010) 
\frac{G_{ii}^{(2),free}}{G_{ii}^{(0),free}}  \; .
\label{RV20}
\end{eqnarray}
Although the statistical significance of the deviation of
$G_V^{(2)}/G_V^{(0)}$ from the free case is marginal, 
the difference is consistent with the behavior
of $G(\tau T)/G^{free} (\tau T)$ close to $\tau T=1/2$ 
shown in Fig.~\ref{fig:GVcont}. At large enough $\tau T$
the latter has a positive slope in $\tau T$, which approaches zero
from above when $\tau T$ reaches $1/2$, and a small negative
curvature that is in the limit of $\tau T$ going to $1/2$ proportional 
to $R^{(2,0)}_V-R^{(2,0)}_{V,free}$, as shown in Eq.~\ref{series}.

We will show in the following that 
this puts stringent bounds on the magnitude 
of any contribution to the vector correlation function that may arise from a 
peak in the vector spectral function at small energies.
We also tried to look at the fourth thermal moment, which could be obtained 
from the curvature of the fits shown in Fig.~\ref{fig:ratios}. However, 
at present our
numerical results do not allow to draw a firm conclusion about its
value.

\section{Analysis of the vector current correlation function}

In the previous section we have presented results for the vector
current correlation function in quenched QCD at $T\simeq 1.45 T_c$. For 
Euclidean times $\tau T\in [0.2,0.5]$ the results could be safely
extrapolated to the continuum limit. As the correlation functions $G_V(\tau T)$
and $G_{ii}(\tau T)$ are closely related to each other, we now concentrate on a 
discussion of the latter.
Let us summarize our basic findings for the $G_{ii}(\tau T)$:
\begin{itemize}
\item[(i)]
The correlation function at $\tau T=1/2$ is about 2\% larger
than the corresponding free field value, $G_{ii} (1/2)/G_{ii}^{free}(1/2)=1.024(8)$
\item[(ii)]
The deviation from the free field value increases with decreasing Euclidean
time. At $\tau T =1/4$ the ratio is $G_{ii} (1/4)/G_{ii}^{free}(1/4)=1.051(7)$
\item[(iii)]
The second moment of the vector spectral function 
deviates from the free field value by about 7\%, 
$G_V^{(2)}/G_V^{(2),free} = G_{ii}^{(2)}/G_{ii}^{(2),free}= 1.067(12)$
\end{itemize}

This suggests that the spectral function of the free theory is a good
starting point for the description of the vector correlator.
We will use the  knowledge about the asymptotic behavior
of the vector spectral function 
at low and high energies, presented in Section~II, to construct 
ans\"atze for the spectral function at high temperature. This is done 
in two steps. In the next subsection we discuss a straightforward
implementation of the low and high energy constraints by using 
a combination of the free continuum spectral function and a 
Breit-Wigner contribution.
Subsequently we will generalize this ansatz and discuss a class of 
spectral functions for which the continuum contribution is cut off at 
some low energy value and present also results from a MEM analysis. 
We find that the most straightforward approach
already gives an excellent description of all our data.

\subsection{Breit-Wigner + continuum ansatz}

We start with the analysis of our results for the vector correlation 
function by using the ansatz,
\begin{eqnarray}
\rho_{00}(\omega) &=& - 2\pi \chi_q  \omega \delta (\omega)  \ ,
\label{fit00} \\
\rho_{ii} (\omega) &=&  
2\chi_q c_{BW}   \frac{\omega \Gamma/2}{ \omega^2+(\Gamma/2)^2}
+ {3 \over 2 \pi} \left( 1 + k \right) 
\; \omega^2  \;\tanh (\omega/4T)   \ .
\label{ansatz}
\end{eqnarray}
This ansatz depends on four temperature dependent parameters; the quark number 
susceptibility $\chi_q(T)$, the strength ($c_{BW}(T)$) and width ($\Gamma (T)$) 
of the Breit-Wigner peak and the parameter $k(T)$ that parametrizes deviations 
from a free spectral function at large energies.
At high temperature and for large energies, $\omega/T \gg 1$, 
we expect to find $k(T)\simeq \alpha_s/\pi$. Note that $k(T)$ will
also depend on $\omega$ and actually will vanish for 
$\omega \rightarrow \infty$ at fixed $T$. We will treat here $k(T)$ as 
a constant and will not take into account any running of $\alpha_s$. 
As will become clear in the following, already this minimal ansatz
provides a good description of current numerical results for the vector
correlation functions $G_{ii} (\tau T)$ and $G_{00} (\tau T)$. In fact, it is 
quite straightforward to use the latter for a determination of $\chi_q(T)$.

The parameters of the Breit-Wigner term can be extracted from fits to 
the vector correlation function. 
As pointed out in Eq.~\ref{relation} the correlation
functions $G_V$ and $G_{ii}$ agree up to an additive constant, the 
quark number susceptibility. It thus suffices to analyze one of them.

With the ansatz for the spectral function given in Eq.~\ref{ansatz} 
and $\widetilde{\Gamma}=\Gamma/T$ we obtain
\begin{equation}
\widetilde{G}_{ii}(\tau T) = \left( 1+k(T) \right) 
\widetilde{G}_{V}^{free}(\tau T) + c_{BW}\widetilde{\chi}_q
F_{BW}(\tau T,\widetilde{\Gamma}) \; ,
\label{ratio_zero_ansatz}
\end{equation}
with
\begin{equation}
F_{BW}(\tau T,\widetilde{\Gamma}) = \frac{\widetilde{\Gamma}}{2\pi} 
\int_{0}^{\infty} {\rm d}\widetilde{\omega}
\frac{\widetilde{\omega}}{(\widetilde{\Gamma}/2)^2+ \widetilde{\omega}^2} 
\frac{\cosh (\widetilde{\omega} (\tau T-1/2))}{\sinh (\widetilde{\omega}/2)} \; .
\label{FBW}
\end{equation}
With this normalization one has 
$\lim_{\widetilde{\Gamma}\rightarrow 0}F_{BW}(\tau T,\widetilde{\Gamma})=1$ 
and
$\lim_{\widetilde{\Gamma}\rightarrow \infty}F_{BW}(\tau T,\widetilde{\Gamma})=0$.

Note that with this ansatz the dependence on the continuum contribution
and thus the dependence on $k(T)$ can easily be eliminated by
taking appropriately weighted differences of $G_{ii}(\tau T)$ at two values of 
Euclidean time.
Similarly we obtain for the mid-point subtracted correlation functions,
introduced in Eq.~\ref{mid-point},
\begin{equation}
\Delta_{V}(\tau T) = 1+k(T)
+ c_{BW}\widetilde{\chi}_q
\frac{F_{BW}(\tau T,\widetilde{\Gamma})-F_{BW}(1/2,\widetilde{\Gamma})}{
\widetilde{G}_V^{free} (\tau T)- \widetilde{G}_V^{free} (1/2)} \; .
\label{ratio_diff_ansatz}
\end{equation}
At $\tau T=1/2$ this fit ansatz yields results for the zeroth and 
second moment of the spectral function as introduced in Eq.~\ref{series}
and Eq.~\ref{mid-point}, 
\begin{eqnarray}
\widetilde{G}_{ii}(1/2) &=& 2\left( 1+k(T) \right)
+  c_{BW} \widetilde{\chi}_q
F^{(0)}_{BW}(\widetilde{\Gamma}) \; , \nonumber \\
\Delta_{V}(1/2) &=& 1+k(T) + c_{BW} \widetilde{\chi}_q
\frac{F^{(2)}_{BW}(\widetilde{\Gamma})}{\widetilde{G}_{V}^{(2),free}} \; ,
\label{Delta12}
\end{eqnarray}
with
\begin{equation}
F^{(2n)}_{BW}(\widetilde{\Gamma}) = \frac{1}{(2n)!}
\frac{\widetilde{\Gamma}}{2\pi}
\int_{0}^{\infty} {\rm d}\widetilde{\omega} \
\frac{\widetilde{\omega}^{2n+1}}{\left( (\widetilde{\Gamma}/2)^2+ \widetilde{\omega}^2 \right)
\sinh (\widetilde{\omega}/2)} \; .
\label{FBWmoment}
\end{equation}
In the limit $\widetilde{\Gamma} \rightarrow 0$ the thermal moments 
$F^{(2n)}_{BW}$ vanish for all $n>0$ and the fit ansatz for 
$\Delta_V (\tau T)$ becomes a constant
which relates to the deviations of the vector correlation function from
the free field correlator at short distances.
For all $\widetilde{\Gamma} >0$, however, the right hand side of 
Eq.~\ref{ratio_diff_ansatz}
is a monotonically increasing function of $\tau T$. We note that even for
our largest lattice this feature is not yet obvious from the data. 
However, it becomes apparent in the
continuum extrapolation of the mid-point subtracted correlation function. 
Having established the monotonic rise of $\Delta_{V}(\tau T)$ is
important for motivating our fit ansatz.

Fitting the continuum extrapolated correlation function $G_{ii} (\tau T)$ 
in the interval $[0.2:0.5]$  
together with the second thermal moment $G_V^{(2)}$ to constrain the fit, we obtain
\begin{equation}
k = 0.0465(30) \; ,\; \widetilde{\Gamma} = 2.235 (75) \; ,\; 
2 c_{BW} \widetilde{\chi}_q/\widetilde{\Gamma} = 1.098 (27) \; .
\label{fitresult}
\end{equation}
This three parameter fit has a $\chi^2/d.o.f. =0.06$ for 12 degrees of freedom. 
The small $\chi^2/d.o.f.$ clearly reflects that even after the continuum
extrapolation data at different distances are strongly correlated.
Nonetheless, the fit provides an excellent description of the data.
\begin{figure}
\begin{center}
\epsfig{file=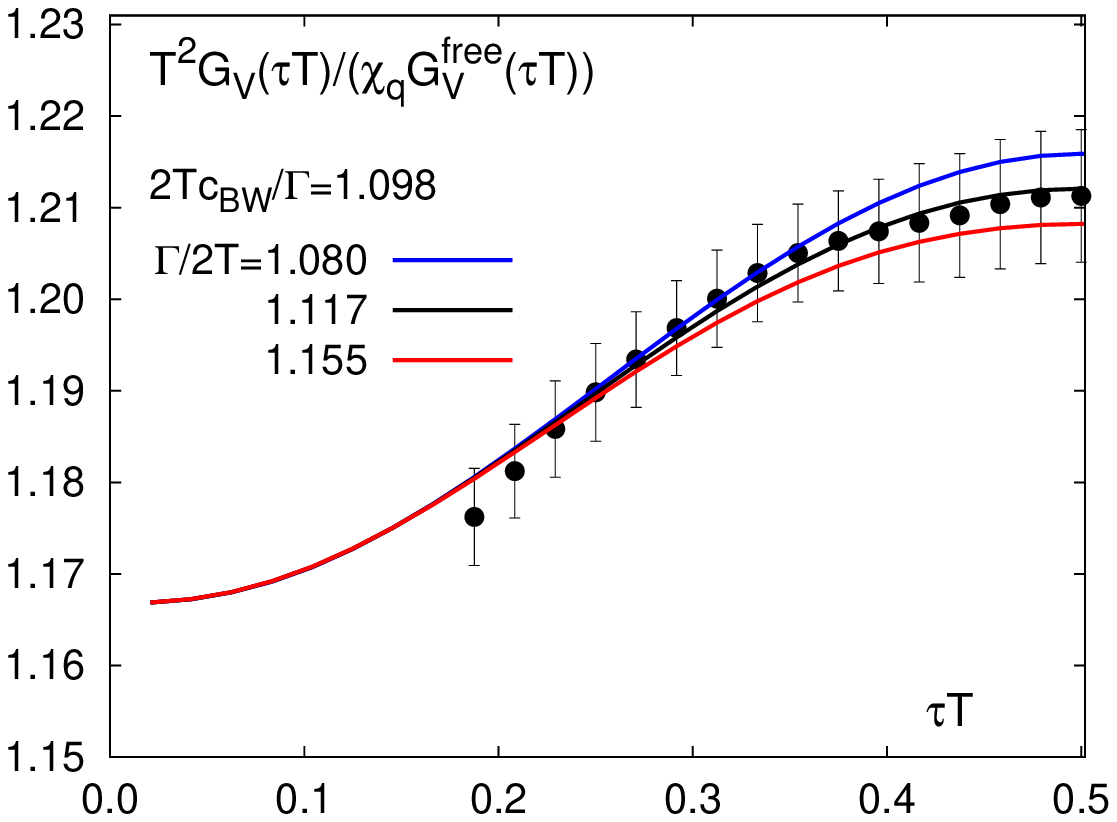,width=77mm}\hspace*{-0.6cm}
\epsfig{file=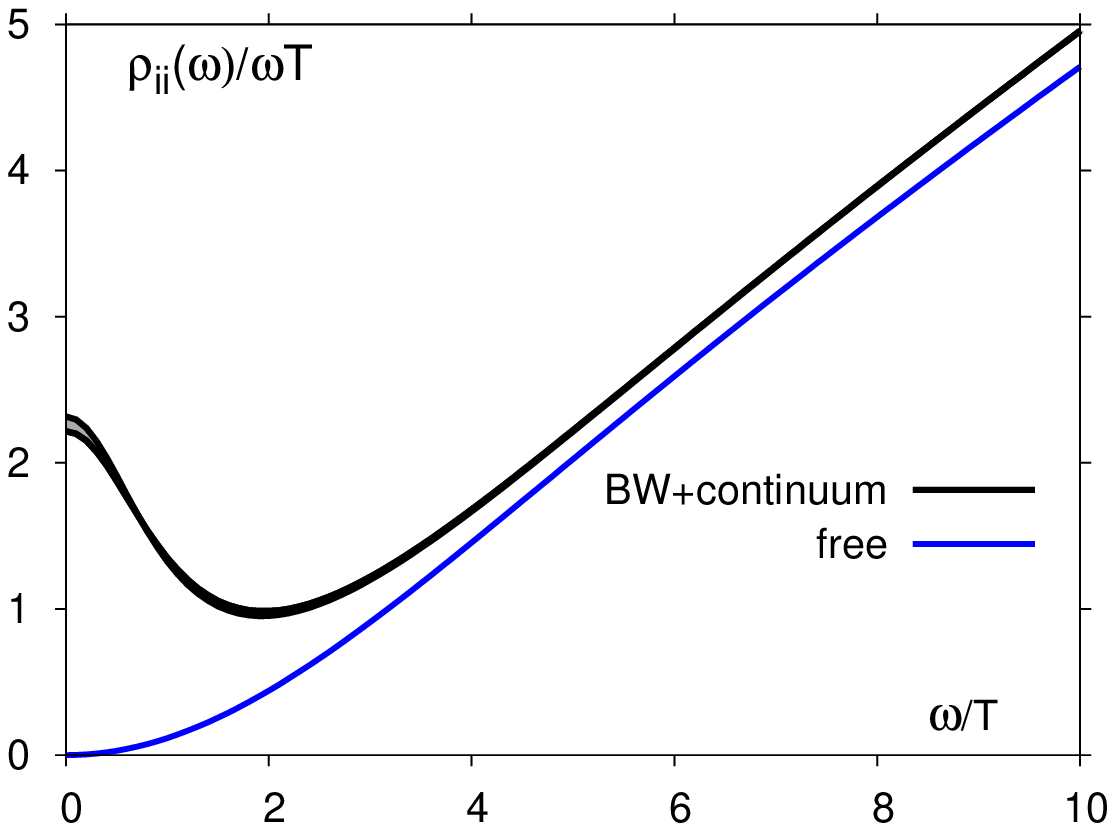,width=77mm}
\end{center}
\caption{\label{fig:fit}
Data for the continuum extrapolation of 
$T^2 G_V(\tau T)/(\chi_q G_V^{free}(\tau T))$ and the fit
result for fixed $c_{BW}/\widetilde{\Gamma}$ and $k(T)$ (left). 
The three curves show the result from a fit in the interval $\tau T\in [0.2:0.5]$
(central) and results obtained by 
varying $\widetilde{\Gamma}$ within its error band. In the right hand
figure we show the spectral function obtained from the fit and compare
with the free spectral function.
}
\end{figure}
To illustrate the sensitivity of our fit to the low energy Breit-Wigner
contribution and its dependence on Euclidean time, we show
the fit to the data for $G_V(\tau T)$ normalized to the 
free vector correlation function and the quark number susceptibility 
in Fig.~\ref{fig:fit}. 
The error band shown in this figure corresponds to the width of the 
Breit-Wigner peak. 
The spectral function obtained from this fit is shown in the right
hand part of the figure. Here also the error band arising from a variation
of the width $\Gamma$ is shown.

It is clear from Fig.~\ref{fig:fit}, that the vector correlation function is
sensitive to the low energy, Breit-Wigner contribution
only for distances $\tau T\gsim 0.25$. 
Taking into account also the value of the second thermal moment, the fits 
to the large distance regime return fit parameters which are well constrained.
As a consequence we obtain a significant result for
the electrical conductivity, which is directly proportional to
the fit parameter $c_{BW}/\widetilde{\Gamma}$,
\begin{equation}
\frac{\sigma}{T} = \frac{C_{em}}{6} \lim_{\omega \rightarrow 0} 
\frac{\rho_{ii}(\omega)}{\omega T} = 
\frac{2 C_{em}}{3} \frac{c_{BW} \widetilde{\chi}_q }{\widetilde{\Gamma}} = 
(0.37\pm 0.01) C_{em} \; , 
\label{electric}
\end{equation}
which (accidentally) is close to the result found in \cite{Aarts07} using
staggered fermions with unrenormalized currents. It is more than an order
of magnitude larger than the electrical conductivity in a 
pion gas \cite{FernandezFraile}.

It should be obvious that this determination of the electrical conductivity
is sensitive to the ansatz made for the spectral function in our 
analysis of the correlation functions. With this simple 
ansatz we obtain good fits of the vector correlation function with a very
small chi-square per degree of freedom. However other ans\"atze 
may provide an equally good description of the current set of data.
We will explore this in the next subsection by generalizing the current ansatz.

We also note that the value determined for the correction to the free field
behavior at large energies $k\simeq 0.05$ at $T\simeq 1.45 T_c$ is quite 
reasonable.  Using the relation to the perturbative result, $k=\alpha_s/\pi$ 
yields for the temperature dependent running coupling 
$g^2(T) =4\pi \alpha_s \simeq 2$ which is in good agreement with other 
determinations of temperature dependent running couplings at high energies
or short distances \cite{Zantow}.

\subsection{Breit-Wigner plus truncated continuum ansatz}

In the ansatz used in the previous subsection the continuum part, {\it i.e.} 
a contribution proportional to the free spectral function, contributes to
$\rho_{ii}(\omega)$ for all $\omega$. Also in the highly non-perturbative low 
momentum region there thus is a contribution proportional to $\omega^3/T$ 
arising from the continuum part. In order to analyze its influence on the 
low energy structure of the spectral function we smoothly truncate the 
continuum contribution at some energy $\omega_0$. We thus replace the spectral 
part of our fit ansatz by
\begin{eqnarray}
\rho_{ii} (\omega) &=&  
2\chi_q c_{BW}   \frac{\omega \Gamma/2}{ \omega^2+(\Gamma/2)^2}
+ {3 \over 2 \pi} \left( 1 + k \right) 
\; \omega^2  \;\tanh (\omega/4T) \Theta(\omega_0,\Delta_\omega)  \nonumber \\
\Theta(\omega_0,\Delta_\omega) &=& 
\left( 1+{\rm e}^{(\omega_0^2-\omega^2)/\omega\Delta_\omega} \right)^{-1}
 \ .
\label{cutansatz}
\end{eqnarray}
In the limit $\Delta_\omega\rightarrow 0$ the function 
$\Theta(\omega_0,\Delta_\omega)$ becomes a $\Theta$-function with 
discontinuity at $\omega_0$. We also used other cut-off 
functions for the continuum contribution, which lead to similar conclusions.
We prefer the above version as it insures that the continuum contribution 
vanishes exponentially at $\omega=0$. This allows, for instance, to replace
the simple free field ansatz, used for the continuum in Eq.~\ref{cutansatz},
by the HTL or further improved spectral functions, which have power-law 
divergences at small $\omega$.

\begin{figure}
\begin{center}
\epsfig{file=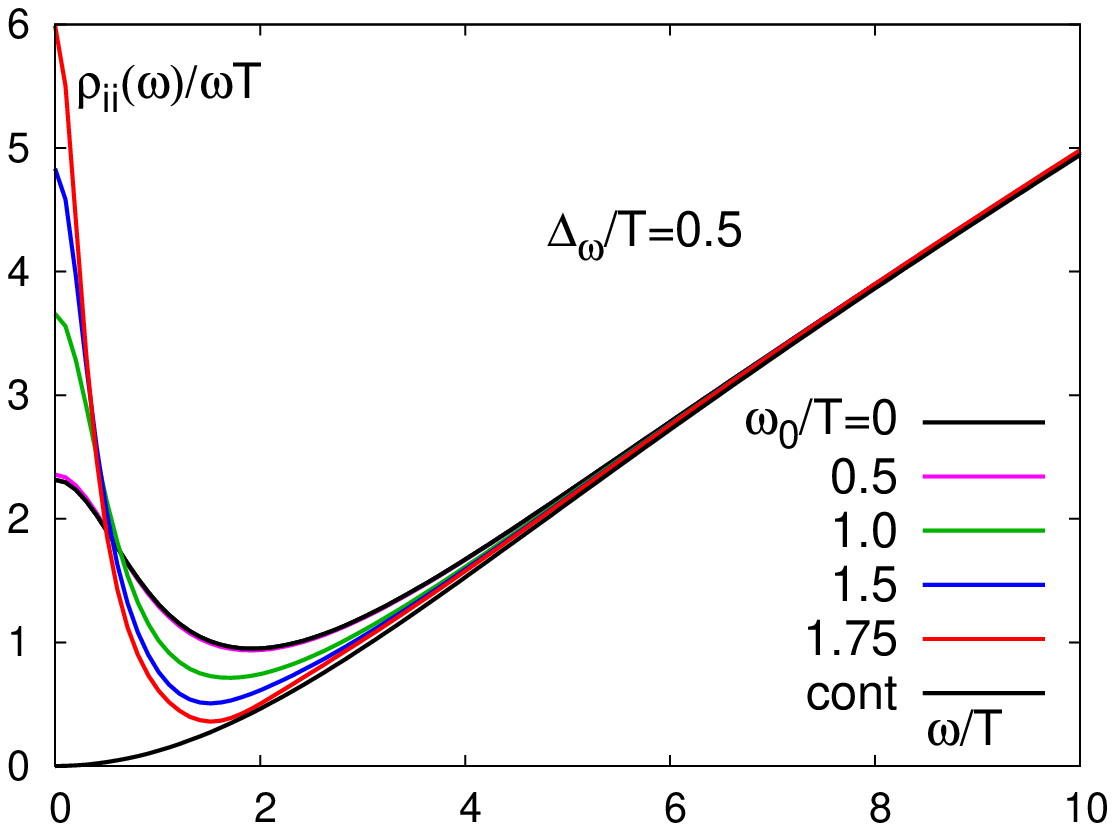,width=77mm}\hspace*{-0.6cm}
\epsfig{file=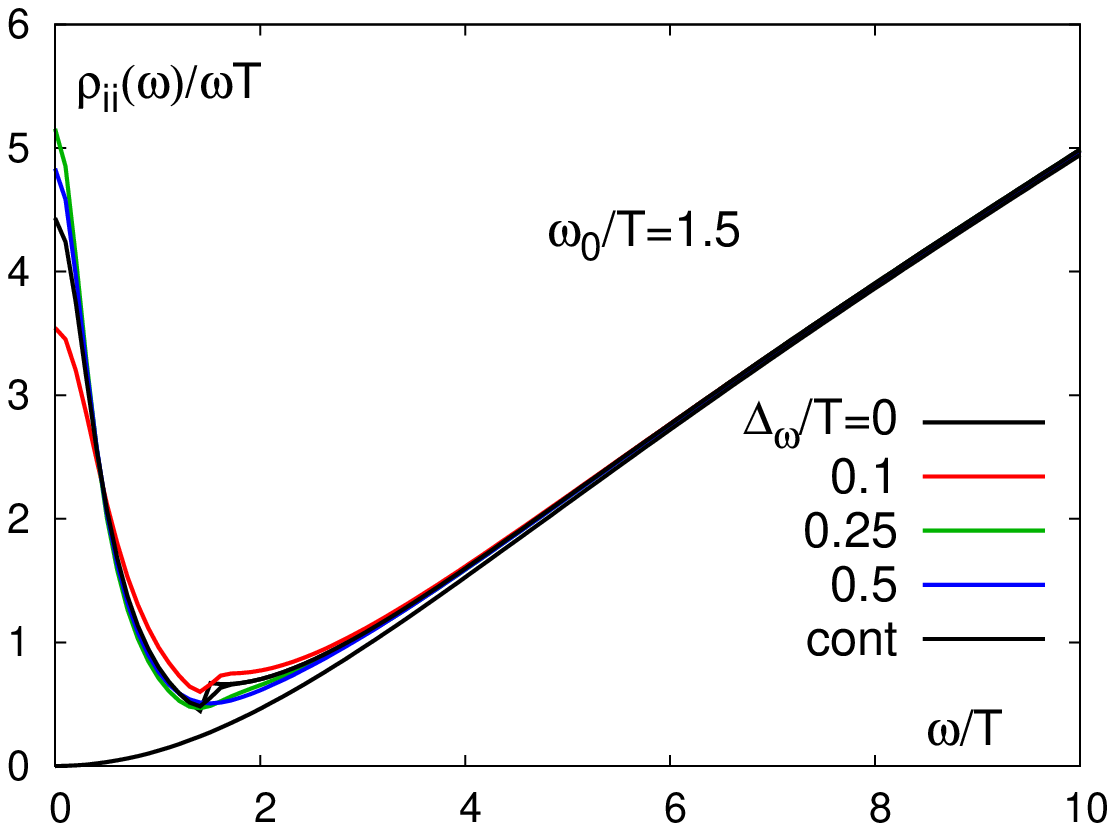,width=77mm}
\end{center}
\caption{\label{fig:cut}
Spectral functions obtained from fits to the vector correlation function
using the ansatz given in Eq.~\protect\ref{cutansatz}. For comparison we
also show only the continuum part of the spectral function.  
The left hand figure shows
results for different values of the cut-off ($\omega_0$) and fixed
width ($\Delta_\omega$).
The right hand figure shows results for
fixed $\omega_0/T=1.5$ and several values of $\Delta_\omega$. The curve labeled
'cont' is the continuum contribution to the fit described in
Eq.~\ref{ratio_zero_ansatz}.
}
\end{figure}

As discussed in the previous section we perform three parameter fits with
$c_{BW}$, $\Gamma$ and $k$ as free parameters for several values of 
$\omega_0$ and $\Delta_\omega$. We find that fits become worse with 
increasing $\omega_0$ and/or increasing $\Delta_\omega$. In both cases
eventually too much of the continuum part at high energies gets 
suppressed. For small values of $\omega_0$ and $\Delta_\omega$ the 
Breit-Wigner term compensates for the continuum contribution that has been 
cut off by increasing the low energy contribution, {\it i.e.} the 
intercept at $\omega =0$ (electrical conductivity) rises. Results from
fits which all lead to $\chi^2/d.o.f.$ smaller than unity are shown in
Fig.~\ref{fig:cut}. 
As $\omega_0$ and $\Delta_\omega$ increase the $\chi^2/d.o.f.$ of the
fits shown in this figure rises from its minimal value of about 0.06,
obtained for $\omega_0/T=\Delta_\omega/T=0$, to unity. All fit parameters
corresponding to the curves shown in Fig.~\ref{fig:cut} are summarized 
in Table~\ref{tab:fit_parameter}

\begin{table}[t]
\begin{center}
\vspace{0.3cm}
\begin{tabular}{|c|c|c|c|c|c|}
\hline
$\omega_0/T$ & $\Delta_\omega/T$ & $2c_{BW}\widetilde{\chi}_q/\widetilde{\Gamma}$ &
$\widetilde{\Gamma}$ & $k$ & $\chi^2/dof$\\
\hline
0.0 & 0.5 & 1.290(46) & 2.091(112) &0.1677(42) & 0.08 \\
0.5 & 0.5 & 1.315(43) & 2.038(114) &0.1683(41) & 0.11 \\
1.0 & 0.5 & 2.039(22) & 1.198(25) &0.1739(4) & 0.19 \\
1.5 & 0.5 & 2.694(19) & 0.866(15) &0.1760(4) & 0.56 \\
1.75 & 0.5 & 3.338(18) & 0.679(15) &0.1774(4) & 1.00 \\
\hline
1.5 & 0.0 & 2.471(20) & 0.947(17) &0.1778(4) & 0.32 \\
1.5 & 0.1 & 1.976(23) & 1.232(27) &0.1741(4) & 0.36 \\
1.5 & 0.25& 2.873(19) & 0.808(13) &0.1773(4) & 0.39 \\
1.5 & 0.5 & 2.694(19) & 0.866(15) &0.1760(4) & 0.56 \\
\hline
\end{tabular}
\end{center}
\caption{Parameters for the fits shown in Fig.~\ref{fig:cut}
left (upper part) and right (lower part). The last column gives
the $\chi^2/dof$ of these fits.
}
\label{tab:fit_parameter}
\end{table}

\begin{figure}
\begin{center}
\epsfig{file=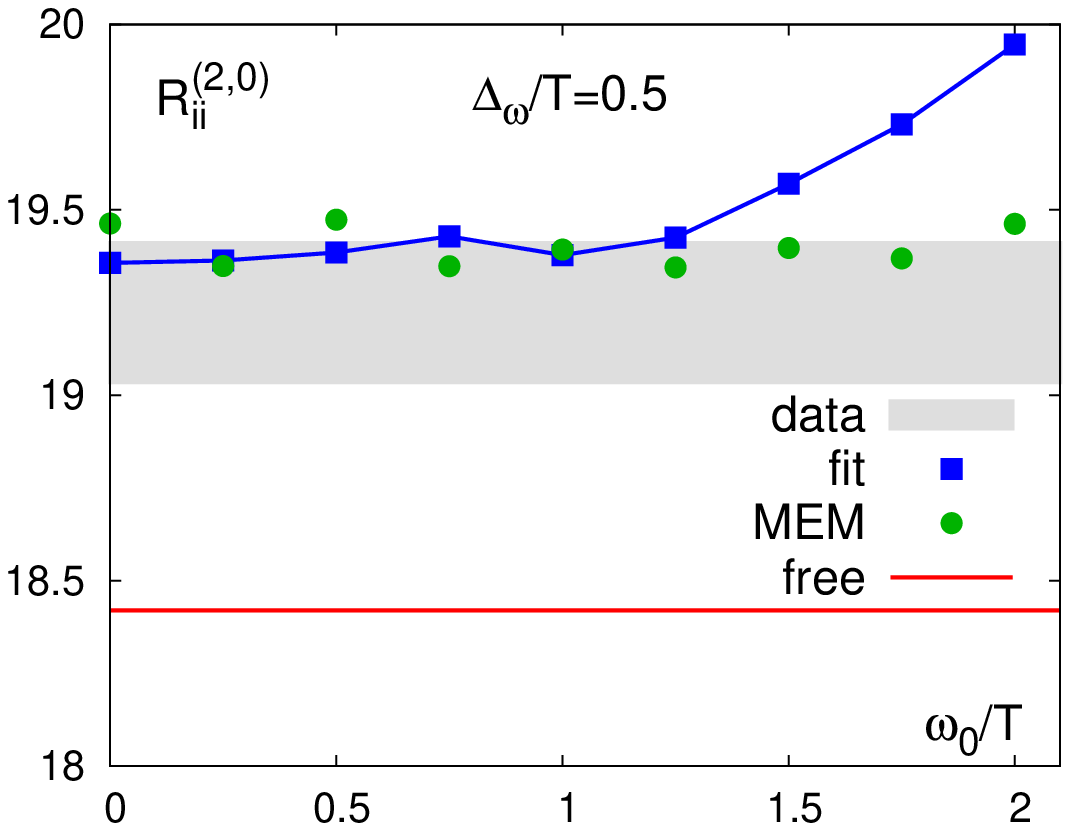,width=88mm}
\end{center}
\caption{\label{fig:moment_fit}
The ratio of second and zeroth thermal moment of the correlation
function $G_{ii}(\tau T)$ obtained from fits with different values for
the continuum cut-off parameter $\omega_0/T$ and fixed $\Delta_\omega$.
Circles show results of a MEM analysis where the fits have been used
as default model.
The band gives the result extracted from the continuum extrapolated
correlation function. The lower curve shows the corresponding free 
field (infinite temperature) value, which is about 5\% smaller than the 
value obtained at $T\simeq 1.45 T_c$.
}
\end{figure}

In particular the second moment of the correlation function normalized
by the correlation function at the mid-point, {\it i.e.} the ratio 
$R_{ii}^{(2,0)}$ introduced in Eq.~\ref{series}, reacts quite sensitive to 
the truncation of the continuum part of the spectral function. This is 
shown in Fig.~\ref{fig:moment_fit}, where we compare the ratio $R_{ii}^{(2,0)}$
extracted from our continuum extrapolated data (error band) with fit results
obtained for different values of $\omega_0$. The dependence on $\Delta_\omega$,
on the other hand, is less pronounced.

We conclude from this analysis that for $\omega/T\gsim (2-4)$ acceptable 
spectral functions should not deviate significantly from the perturbative 
or free field like behavior. Moreover, the structure of the spectral 
function in the energy range $\omega/T \lsim 2$ is sensitive to the 
form of the fit ansatz. Within the class of functions analyzed here the
fits do, however, favor a small value for the cut-off $\omega_0$ and 
a small value for the intercept of $\rho_{ii}(\omega)/\omega$
at $\omega =0$. 

These findings are consistent with the default model dependence of a MEM 
analysis which we will briefly discuss in the next subsection.

\subsection{Analysis of vector correlation functions using the Maximum Entropy
Method}

So far we did not make use of the Maximum Entropy Method that has been used
in most other lattice studies of the spectral function in the vector channel. 
Here we want to discuss to what extent the analysis presented
in the previous subsections can be reproduced in a MEM analysis, or whether a
MEM analysis may improve over the result obtained with an ansatz for the
spectral function. We performed a MEM analysis of the renormalized 
vector correlation function on our finest lattice
using the modified kernel which has been introduced in \cite{Aarts07} and
further refined in \cite{engels10}.

\begin{figure}
\begin{center}
\epsfig{file=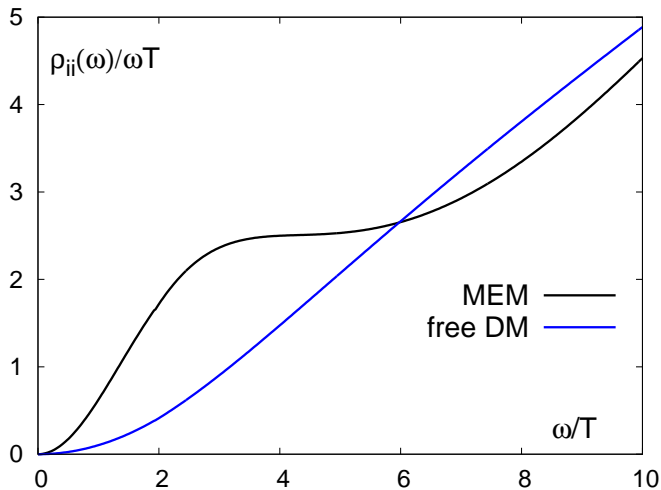,width=98mm}
\end{center}
\caption{\label{fig:freespf}
Spectral functions obtained from a maximum entropy analysis using
the free spectral function as default model. 
}
\end{figure}

To start the MEM analysis of a given meson correlation function we need to
specify a default model that incorporates all prior knowledge about the
spectral function we want to determine. 
Before presenting the analysis based on default models related to the
fits discussed in the previous subsections we performed a MEM analysis
in analogy to what we had done earlier \cite{our_dilepton}.
We used as default model the
free spectral function, {\it i.e.} an ansatz that does not include 
a contribution linear in $\omega$ at small energies but rather 
 $\rho_{ii}^{DM}(\omega)/\omega \sim \omega^2/T$. This default model
together with the  resulting output spectral function is shown in 
Fig.~\ref{fig:freespf}. Apparently this ansatz also forces 
$\rho^{mem}(\omega)/\omega$ to vanish at $\omega =0$. 
In particular, it leads to a suppression of the dilepton rate at small
energies.
The small contribution
at small energies is compensated by large contributions at  
$\omega \sim (2-4)T$. This feature has been observed also in our first analysis
of vector spectral functions \cite{our_dilepton}.  

To allow a MEM analysis to find a contribution $\rho(\omega)\sim \omega$
at the origin this possibility needs to be offered by the default 
model\footnote{On the other hand, in correlation functions for quantum number 
channels where no transport contribution is present, MEM is flexible enough to
suppress or eliminate completely a contribution $\rho(\lambda)\sim\lambda$ 
included in the default model. This is, for instance, the case in the 
pseudo-scalar channel.}
It thus seems to be appropriate
to use the result of the analysis presented in the previous subsections as 
default model in a MEM analysis. It definitely summarizes the best knowledge
we have at present about the spectral function in the vector channel.
Using the un-modified version of the Breit-Wigner plus continuum ansatz, 
Eq.~\ref{ansatz}, as default model changes results only little 
and leads to an output spectral function that is compatible with the input 
form. It also reproduces the result obtained for the electrical conductivity 
given in Eq.~\ref{electric}. 

\begin{figure}
\begin{center}
\epsfig{file=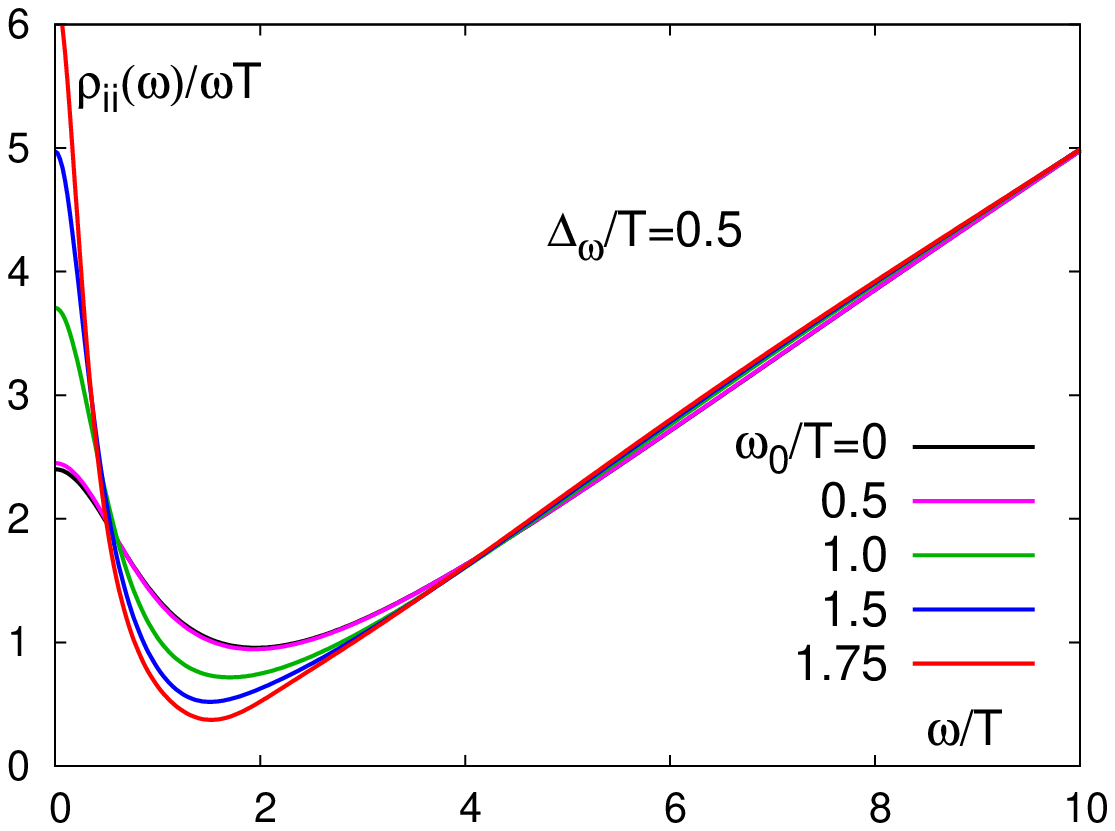,width=78mm}\hspace*{-0.8cm}
\epsfig{file=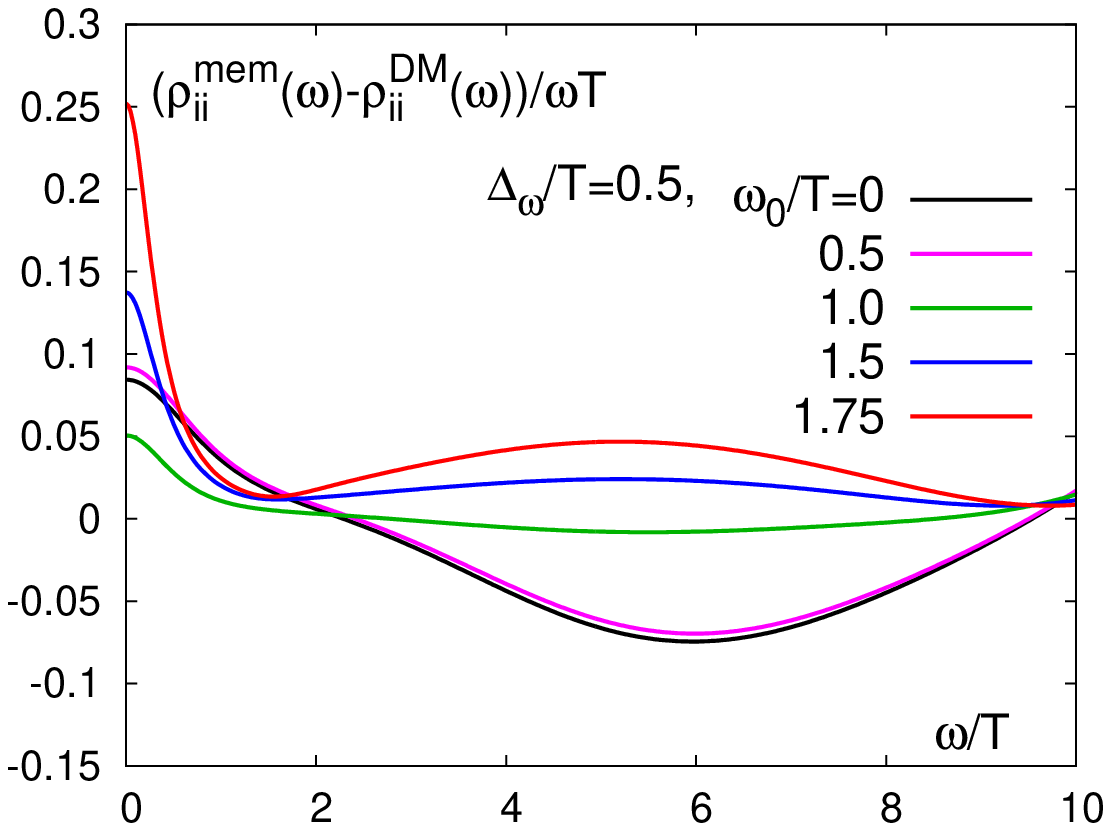,width=78mm}
\end{center}
\caption{\label{fig:mem}
Spectral functions obtained from a maximum entropy analysis using
for the default model the spectral functions shown in
Fig.~\ref{fig:cut}(left). The right hand figure shows the difference
between the output spectral function obtained from the MEM analysis and
the input spectral function used in each case.
}
\end{figure}

In order to judge the stability of the spectral function obtained with our fit
ansatz, we used a 
class of default models of the form given in Eq.~\ref{cutansatz}.
We performed a MEM analysis using 
as input default models the spectral
functions shown in Fig.~\ref{fig:cut}. 
We controlled the error on the output
spectral functions by performing the MEM analysis on jackknife blocks.
In this way we find that the error of the output spectral functions never
is larger than 10\% for small energies and is below 5\% for $\omega/T\gsim 8$.
In Fig.~\ref{fig:mem}(left) we show results for spectral functions
obtained in a MEM analysis by using a set
of default models with varying $\omega_0$ and fixed $\Delta_\omega$.
The right hand figure shows the difference between the
input default models and the output spectral functions.
We note that these differences correspond to changes in the spectral
functions which are generally smaller than 5\%. They increase
in particular for small values of $\omega$ as the $\chi^2/d.o.f.$ of the 
input default model gets worse. We also note that the MEM analysis reproduces
the calculated ratio of thermal moments as well or even better than the
fits used as a default model, although the value of the thermal moments 
itself did not directly enter the MEM analysis. 

\section{Electrical conductivity, thermal dilepton and photon rates}

Here we want to discuss the consequences of our results for thermal
dilepton rates as well as the electrical conductivity and the related
soft photon rate.

\begin{figure}
\begin{center}
\epsfig{file=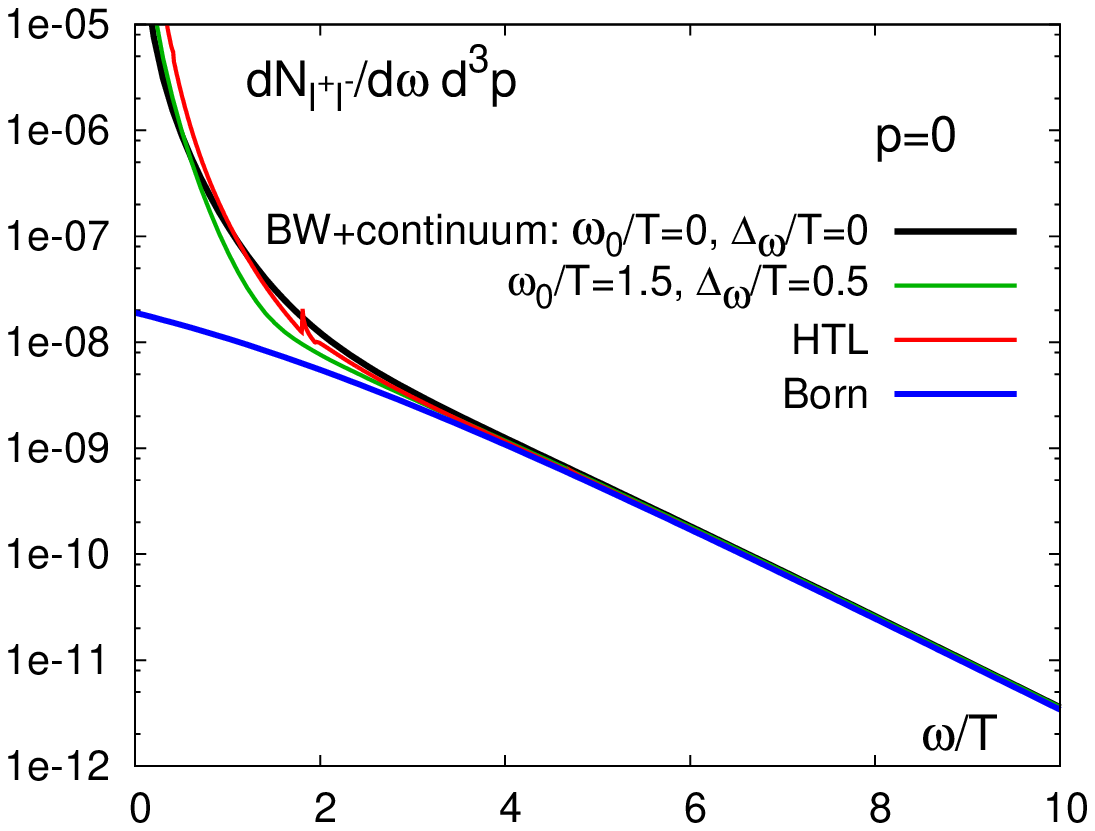,width=78mm}\hspace*{-0.8cm}
\epsfig{file=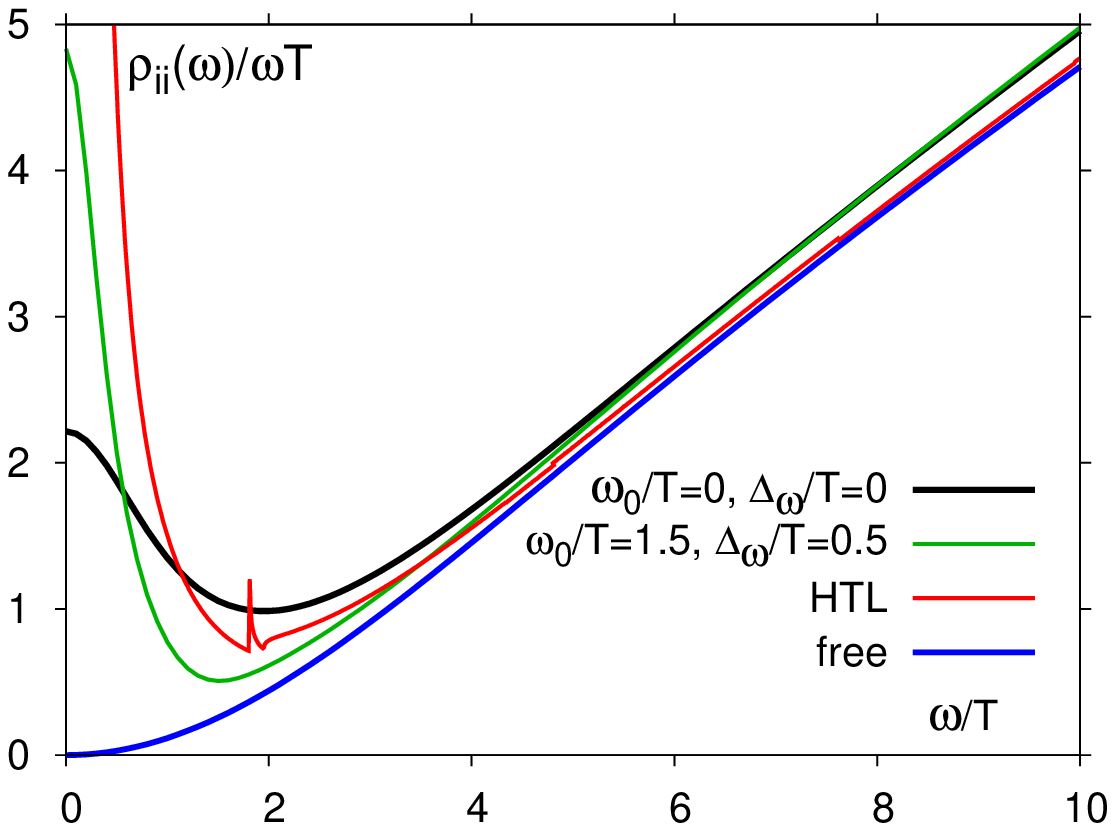,width=78mm}
\end{center}
\caption{\label{fig:dilepton}
Thermal dilepton rate in 2-flavor QCD (left). Shown are results from
fits without a cut-off on the continuum contribution ($\omega_0/T=0$) and
with the largest cut-off tolerable in our fit ansatz ($\omega_0/T=1.5$).
The HTL curve is for a thermal quark mass $m_T/T=1$ and the 
Born rate is obtained by using the free spectral function. The right hand 
part of the figure shows the spectral functions that entered the calculation 
of the dilepton rate.
}
\end{figure}

In Fig.~\ref{fig:dilepton} we show the thermal dilepton rate calculated 
from Eq.~\ref{rate} for two massless ($u,\ d$) flavors. We use the results
obtained with our Breit-Wigner plus continuum fit ansatz, Eq.~\ref{ansatz},
as well as results obtained with a truncated continuum term. For the
latter we use the case, $\omega_0/T=1.5$, $\Delta_\omega/T =0.5$,
which gave a $\chi^2/d.o.f$ of about 1. These results are compared to 
a dilepton spectrum calculated within the hard thermal loop approximation
\cite{Braaten} using a thermal quark mass $m_T/T=1$. Obviously the results 
are in good agreement for all $\omega/T\gsim 2$. For $1\lsim \omega/T\lsim 2$ 
differences between the HTL spectral function and  
our numerical results is about a factor two, which also is the intrinsic
uncertainty in our spectral analysis. At energies $\omega/T\lsim 1$ 
the HTL results grow too rapidly, as is well known.

In the limit $\omega \rightarrow 0$ the results for $\rho_{ii}(\omega)/\omega$,
and thus also for the electrical conductivity, are sensitive to
the choice of fit ansatz. Within the class of ans\"atze used by us a small
value of $\rho_{ii}(\omega)/\omega$ seems to be favored. Our current analysis
suggests,
\begin{equation}
2 \ \lsim \ \lim_{\omega \rightarrow 0} \frac{\rho_{ii}(\omega)}{\omega T} \ 
\lsim \ 6 \quad {\rm at} \quad T\simeq 1.45\ T_c\; .
\label{range_sigma}
\end{equation}
This translates into an estimate for the electrical conductivity
\begin{equation}
1/3 \ \lsim \ \frac{1}{C_{em}}
\frac{\sigma}{T} \ \lsim \ 1 \quad {\rm at} \quad T\simeq 1.45\ T_c \; .
\label{range}
\end{equation}
Using Eq.~\ref{photon} this yields for the zero energy limit of the thermal 
photon 
rate\footnote{Here we used $T_c\simeq 165$~MeV. 
This is a value relevant 
for QCD with 2 light quarks rather than the critical temperature for a pure 
$SU(3)$ gauge theory.},
\begin{equation}
\lim_{\omega \rightarrow 0} \omega \frac{{\rm d} R_\gamma}{{\rm d}^3p} =
\left( 0.0004 \ -\ 0.0013 \right) T_c^2 \simeq (1-3)\cdot 10^{-5}\ {\rm GeV}^2
\quad {\rm at} \quad T\simeq 1.45\ T_c  \ .
\label{softphoton2}
\end{equation}

\section{Conclusions}

At a fixed value of the 
temperature, $T \simeq 1.45 T_c$, we 
have performed a detailed analysis of vector correlation functions
in the high temperature phase of quenched QCD. A systematic analysis
at different values of the lattice cut-off 
combined with an analysis of finite volume and 
quark mass effects allowed us to extract the vector correlation 
function in the continuum limit for a large interval of Euclidean
times, $0.2 \le \tau T\le 0.5$. In this interval the correlation 
function has been determined to better than 1\% accuracy. Furthermore,
we determined its curvature at the mid-point of the finite temperature
Euclidean time interval, $\tau T=1/2$.

We analyzed the continuum extrapolated vector correlation functions using 
several fit ans\"atze that differ
in their low momentum structure. We find that the vector correlation 
function is best fitted by a simple ansatz that is proportional to a
free spectral function plus a Breit-Wigner term. Already this simple
three parameter ansatz yields a good description of the correlator with
a small $\chi^2/d.o.f.$. Other fits with a $\chi^2/d.o.f. \le 1$ are possible
and suggest that the low energy structure of the spectral function
currently has a systematic uncertainty of about a factor two. Some generic
features of the spectral function are, however, robust. For energies
$\omega/T \gsim (2-4)$ the spectral function is close to the free 
form. At present we used an ansatz proportional to the free spectral function
to parametrize the spectral function in this regime. However,
the hard thermal loop spectral function should give a good description at
these energies and can be incorporated in our ansatz, which cuts off the 
small energy continuum contributions. 

For small energies, $\omega/T \lsim (1-2)$,
the spectral function is significantly larger than the free result,
but smaller than the HTL spectral function, which diverges at small
energies; at
energies $\omega/T\simeq 1$ the thermal dilepton rate is about an order of 
magnitude larger than the leading order Born rate. In order to analyze 
quantitatively to what extent this 
enhancement can account for the experimentally observed enhancement of 
dilepton rates at low energies \cite{Phenix,Ceres} we will need results on 
the spectral function at temperatures closer to the transition temperature
as well as knowledge on its momentum dependence. With this a complete
analysis of dilepton rates that takes into account the hydrodynamic
expansion of dense matter created in heavy ion collision will become
possible \cite{Rapp}.

\section*{Acknowledgments}
\label{ackn}
This work has been supported in part by contracts DE-AC02-98CH10886
with the U.S. Department of Energy, the BMBF under grant 06BI401, the
Gesellschaft f\"ur Schwerionenforschung under grant BILAER, the Extreme
Matter Institute under grant HA216/EMMI and the Deutsche
Forschungsgemeinschaft under grant GRK 881.
Numerical simulations have been performed on the BlueGene/P at the New York
Center for Computational Sciences (NYCCS) which is supported by the U.S.
Department of Energy and by the State of New York
and the John von Neumann Supercomputing Center (NIC) at
FZ-J\"ulich, Germany.

\end{document}